\documentclass[prl,twocolumn,superscriptaddress,amsmath,amssymb,floatfix]{revtex4-2}
\usepackage[utf8]{inputenc}
\usepackage[english]{babel}
\usepackage{graphicx}
\usepackage{xcolor}
  \usepackage{hyperref}
  \hypersetup{
    colorlinks=true,
    allcolors=black,
    hypertexnames=false,   
    plainpages=false,
  }
\usepackage{bibunits}
\defaultbibliographystyle{apsrev4-2}
\defaultbibliography{just-a-clock}

\begin{document}

\title{Conditioning as a route to stereotyped behavior in growing populations}
\author{Riccardo Ravasio}
\affiliation{Department of Physics, University of Chicago, Chicago, IL 60637}

\author{Kabir Husain}
\affiliation{Department of Physics, University of Chicago, Chicago, IL 60637}
\affiliation{Department of Physics and Astronomy, and Laboratory for Molecular Cell Biology, University College London, United Kingdom}

\author{Constantine G. Evans}
\affiliation{Hamilton Institute, Maynooth University, Maynooth, Ireland}

\author{Rob Phillips}
\affiliation{Department of Physics, Caltech, Pasadena, CA 91125}

\author{Marco Ribezzi-Crivellari}
\affiliation{Laboratoire de Biochimie, Chimie Biologie et Innovation, ESPCI Paris, Universit\'e PSL, Paris, France}

\author{Jack W. Szostak}
\affiliation{Howard Hughes Medical Institute, Department of Chemistry, The University of Chicago, Chicago, IL, USA}

\author{Arvind Murugan}
\affiliation{Department of Physics, University of Chicago, Chicago, IL 60637}
\begin{abstract}
Biological systems perform complex multi-step processes in a reproducible way despite underlying stochasticity. The standard explanation is micromanagement by molecular machinery that recognizes and corrects specific errors. Here we study conditioning, a qualitatively different strategy in which attempts failing a coarse criterion are destroyed and do not leave a physical record. The surviving, i.e., conditioned, ensemble is narrower and therefore more ordered. We model conditioning through stochastic resets in a ``socks-before-shoes'' model of a growing population, where $n$ actions must be completed in any order to replicate and any replication attempt not finished by a threshold time is discarded. We find that resets impose hierarchical temporal ordering of the $n$ actions without microscopic control over which action happens when. When disorder carries a sufficient time penalty, this ordering is free: the fastest-growing population is automatically the most ordered, with no direct selection for order required. Save points, at which verified progress is preserved across resets, allow conditioning to scale to complex multi-step processes. Conditioning provides a minimal route to reliable behavior, requiring only a clock rather than molecular machinery that recognizes specific errors. For the right class of processes, it pays for itself.
\end{abstract}
\maketitle

\begin{bibunit}

Biological processes at many scales navigate high-dimensional spaces in a stereotyped manner, i.e., reliably performing their constituent steps in the same sequence.
A genome is copied base by base across templates with $\sim 10^6$ bases; a ribosome is assembled component by component from dozens of subunits; a developing embryo progresses through a sequence of coordinated cell divisions and differentiations. The underlying dynamics are stochastic due to thermal fluctuations, diffusion-limited encounters, variable copy numbers and other factors. One might expect the resulting trajectories to wander widely, producing highly variable outcomes. Yet biological systems often show ordered and reproducible behavior, with far less trajectory-to-trajectory variation than the raw stochasticity and dimensionality would suggest. How do trajectories become so stereotyped?

The standard picture invokes specific error correction: molecular machinery that recognizes when the system has deviated from the correct trajectory and acts to fix the specific deviation. Such mechanisms do exist; mismatch repair enzymes recognize specific base-pair mismatches in newly copied DNA \cite{kunkel_dna_2005}, and the adaptive immune system generates receptors that bind particular molecular targets \cite{mckeithan_kinetic_1995,mayer_diversity_2016}. However, extending this strategy to every biological process that shows ordered trajectories faces a high-dimensional recognition problem. The space of possible deviations is vast, and the correction machinery must distinguish correct from incorrect states at each step. Such a naïve model of chaperones, for example, would require them to recognize every kind of misfold due to every kind of mutation across each of their many client proteins, an enormous molecular recognition problem.

Here, we explore a qualitatively different route to achieving stereotyped behavior that requires no specific error recognition: conditioning \cite{jaynes_information_1957,hachmo_conditional_2023}. Biological systems constitutively destroy and turn over their own components \cite{reddien_purpose_2024,kondev_exploratory_2026}: stalled ribosome assemblies are disassembled and the parts re-used \cite{brandman_ribosome_2016}, polymerases that misincorporate a base backtrack before extending, and misfolded proteins are degraded before they can act on substrates \cite{hartl_molecular_2011,balchin_invivo_2016}. These are instances of exploratory dynamics \cite{kondev_exploratory_2026}, in which the physical record of failures is destroyed and the system tries again. Previous work has focused on the speed of such processes \cite{ravasio_evolution_2026}; our interest is in the structure of the surviving trajectory ensemble. The surviving ensemble is a conditional distribution $P[x(t) \mid C]$, where $C$ is the implicit criterion of the turnover mechanism \cite{hachmo_conditional_2023}: this distribution assigns zero weight to trajectories that fail $C$ and renormalizes probability over those that succeed. 

\begin{figure}
  \includegraphics[width=\columnwidth]{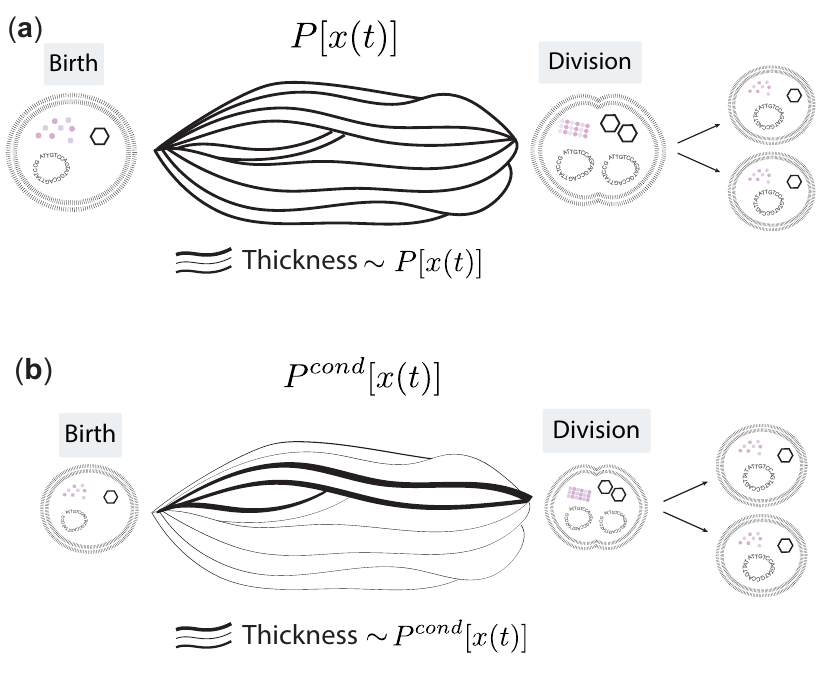}
  \caption{\textbf{Conditioning the trajectory ensemble reduces trajectory entropy.} (a) A self-replicating system traverses stochastic trajectories $x(t)$ from Birth to Division. In the unconditioned ensemble $P[x(t)]$, all trajectories are equally likely (equal line thickness). (b) In physical systems, conditioning arises because trajectories failing a condition $C$ are destroyed before they can leave a record (e.g., incomplete structures are dismantled). The observable ensemble is therefore the conditional distribution $P[x(t) \mid C]$, which concentrates probability on trajectories satisfying $C$ and has lower trajectory entropy: $H[P[\,\cdot\mid C]] \leq H[P]$ (line thickness indicates path weight).}
  \label{fig:concept}
\end{figure}

We model such turnover of replication attempts through stochastic resets \cite{evans_diffusion_2011,roldan_stochastic_2016,evans_stochastic_2020,pal_inspection_2022,alston_boosting_2025}: any attempt not completed by time $T_r$ is aborted and restarted. Unlike classical error correction, which requires recognizing specific deviations (a Maxwell demon that must read microscopic states), a reset needs only a clock. In the reset-conditioned ensembles studied below, the conditioned ensemble is narrower than the unconditioned one; see Fig.~\ref{fig:concept} for an illustration of this trajectory-entropy reduction by conditioning. As Jaynes emphasized \cite{jaynes_information_1957,hachmo_conditional_2023,kondev_exploratory_2026}, such conditioning can make the surviving ensemble appear directed, as though an effective force were guiding trajectories toward completion in a specific way, even when the underlying dynamics are entirely unchanged. 

Conditioning has a potential cost: aborting events necessarily wastes time, which, in the context of a self-replicating system, could reduce its growth rate and thereby its fitness. Whether conditioning is favored depends on whether the benefit of more ordered behavior outweighs this penalty. We study this tradeoff in a ``socks-before-shoes'' model where individuals in a growing population must complete $n$ actions to replicate and any replication attempt not finished by a threshold time is discarded (i.e., is reset). Resets produce hierarchical ordering of actions without any microscopic control over sequencing, and save points at which progress is preserved allow this strategy to extend to complex processes. We find that in parameter regimes where disordered trajectories are rare but slow, producing heavy-tailed completion times, conditioning improves both growth rate and order simultaneously. Conditioning mechanisms can therefore evolve spontaneously, even without any selection for order. Using experimental measurements of stalling times across diverse biochemical processes, we predict quantitatively when this should occur for kinetic proofreading-like mechanisms.

\section{Theory}
Consider a self-replicating system that traverses a stochastic trajectory $x(t)$ from an initial state (Birth) to a final state (Division) in order to replicate, as illustrated in Fig.~\ref{fig:concept}. The space of possible trajectories is high-dimensional, and the distribution $P[x(t)]$ over trajectories characterizes the stochastic dynamics fully. Each trajectory $x(t)$ has a completion time $T[x(t)]$; $P(T)$ is the distribution over completion times associated with $P[x(t)]$.
In an exponentially growing population, fitness is the long-term growth rate $f_\text{growth}$ and is set by lineage amplification rather than by the mean replication time \cite{jafarpour_bridging_2018}. We define the moment generating function $\tilde{P}(w)$ of $P(T)$, $\tilde{P}(w) = \int_0^\infty e^{wT} P(T)\, dT$. Then, the fitness $f_\text{growth}$ for a distribution $P(T)$ is implicitly given by the solution to the Euler-Lotka equation
\begin{equation}
2 \tilde{P}(-f_\text{growth}) = 1,
\label{eq:1}
\end{equation}
where the factor $e^{-f_\text{growth}T}$ upweights faster replication events in setting $f_\text{growth}$ and the factor $2$ accounts for symmetric division of a cell into two daughters. The Euler--Lotka equation \cite{sharpe_problem_1911,jafarpour_bridging_2018} weights replication events by their time-to-division: in an exponentially growing background, earlier divisions contribute more strongly than later ones; see Supplementary Information for a derivation.

Conditioning on completion before a reset time $T_r$, i.e., restarting any attempt that has not completed by $T_r$ and removing the record of such trajectories, modifies both the observed trajectory ensemble and the fitness. The conditioned trajectory distribution is $P^\text{cond}[x(t)] = P[x(t) \mid T[x] < T_r]$: trajectories that do not complete by $T_r$ are assigned zero weight and the rest are renormalized. The acceptance fraction $a(T_r) = \int_0^{T_r} P(T)\, dT$, computable from the induced distribution of completion times $P(T)$, is the probability that a trajectory completes before $T_r$. For the models and parameter regimes analyzed here, the conditioned distribution $P^\text{cond}[x(t)]$ has lower Shannon entropy than $P[x(t)]$: temporal conditioning preferentially removes the slow, disordered trajectories that dominate the entropy of the unconditioned ensemble. We quantify this stereotyping through the Shannon entropy difference
\begin{equation}
\Delta S(T_r) \equiv H[P] - H[P^\text{cond}],
\label{eq:2}
\end{equation}
where $H[P] = -\int P[x(t)] \ln P[x(t)]\, \mathcal{D}x(t)$.

Conditioning has two effects on fitness: it modifies the growth rate $f_\text{growth}(T_r)$ and it reduces trajectory entropy by $\Delta S(T_r)$. We combine these two effects into a total fitness $F$ to study the synergy between these two terms,
\begin{equation}
\ln F(T_r) = \ln f_\text{growth}(T_r) + s\,\Delta S(T_r),
\label{eq:3}
\end{equation}
where $s \in [0,1)$ serves as the selection coefficient for trajectory entropy reduction; see Section~\ref{sec:SI-s-range}. At $s = 0$, fitness reduces to pure growth-rate selection; as $s$ increases, stereotyping becomes increasingly important. Note that the entropy term $\Delta S$ rewards low trajectory entropy rather than correctness per se. In many biophysical contexts, errors correlate with slowness (as with stalling \cite{rajamani_effect_2010}, kinetic traps \cite{jhaveri_discovering_2024}), and so the only trajectories that survive temporal conditioning are those that are error-free \cite{ravasio_evolution_2026}. Consequently, the second term implicitly selects against errors as well.

Stereotyped behavior is a precondition for function and can sometimes be functional in itself. Cell-surface glycans enable immune recognition precisely because their synthesis is reproducible. They serve as molecular identity markers; the particular sugar pattern is evolutionarily arbitrary, but it must be reproducible so the immune system can distinguish self from non-self \cite{varki_biological_2017}. Given equal alternatives, foraging ants converge on a single trail because a consistently reinforced trail is followable, while a diffuse cloud of pheromone provides no directional information \cite{camazine_self-organization_2001}. A mirror-image biochemistry built on D-amino acids would be equally functional, but a mixture of L and D would prevent stable protein folding \cite{blackmond_origin_2010}. Multi-subunit self-assembly, such as viral capsids and microtubules, requires uniform subunit conformations with variable subunits producing amorphous aggregates \cite{perlmutter_mechanisms_2015}. Downstream processes can be built around any consistent input, but not around a variable one. We therefore take $\Delta S$ as a general measure of stereotyped behavior. We will later show that selection on $\Delta S$ is not necessary for the evolution of conditioning.

When resets occur at a fixed reset time $T_r$, each replication attempt either completes before the deadline, with probability $a(T_r)$, or fails to complete and is aborted and restarted, with probability $1-a(T_r)$. The total time to produce a successful offspring is the sum of time spent on failed replication attempts, plus one final successful replication attempt \cite{reuveni_optimal_2016,pal_first_2017}. Averaging over failed replication attempts and using the Euler-Lotka equation, we obtain a relationship between the growth rate $f_\text{growth}$ and the reset time $T_r$
\begin{equation}
2\int_{0}^{T_r} e^{-f_\text{growth}T}\,P(T)\,dT \;=\; 1 - \bigl(1-a(T_r)\bigr)e^{-f_\text{growth}T_r},
\end{equation}
where the integration is now restricted to $[0, T_r]$ because only trajectories completing before the reset contribute, and the right-hand side accounts for the geometric series of failed attempts aborted at $T_r$. This determines $f_\text{growth}(T_r)$; see Supplementary Information for details. The optimal reset time $T_r^*(s)$ is defined as the value maximizing the fitness $F(T_r)$ in Eq.~\ref{eq:3}, balancing the marginal decrease in growth rate against the marginal increase in entropy reduction:

\begin{equation}
\frac{d\ln F}{dT_r}\bigg|_{T_r^*} = \frac{d\ln f_\text{growth}}{dT_r}\bigg|_{T_r^*} + s\,\frac{d\Delta S}{dT_r}\bigg|_{T_r^*} = 0.
\label{eq:4}
\end{equation}
When $s = 0$ the optimum is governed by growth rate alone; as $s$ grows, the stereotyping benefit competes with the growth-rate cost, generally shifting the optimum toward smaller $T_r$. The behavior of $F(T_r)$ depends on the full trajectory distribution $P[x(t)]$: through the induced distribution of completion times $P(T)$ for the growth rate and through the trajectory entropy for $\Delta S$. We study this dependence first in a mechanistic model, then in terms of $P(T)$ alone for a phenomenological survey.

\section{The socks-before-shoes model}
\begin{figure}[h!]
  \includegraphics[width=\columnwidth]{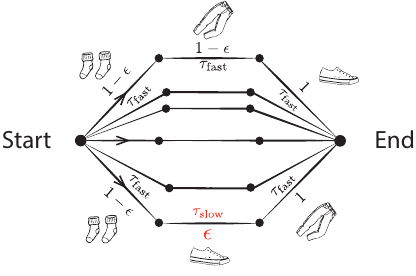}
  \caption{A system must complete $n$ actions in some sequence; correctly ordered actions take time $\tau_\text{fast}$ while misordered actions incur an exponentially distributed slow time cost with mean $\tau_\text{slow} = \eta_\text{stall}\,\tau_\text{fast}$, where $\eta_\text{stall} > 1$ is the dimensionless \emph{stalling factor}. The canonical next action among those remaining is taken with probability $(1-\epsilon)$ and all other remaining actions are taken with uniform probability adding up to $\epsilon$. In the socks-before-shoes example sketched here, the ``shoes'' action needs to be done last, hence will pay the $\tau_\text{slow}$ cost if done before the ``pants'' action.}
  \label{fig:shoessocks_sketch}
\end{figure}

Consider a cell that must complete $n$ actions to replicate, ideally in the canonical order $i = 1,2,3,\ldots n$. But the system frequently errs, performing actions out of order, and such misordered actions are slower, as illustrated in Fig.~\ref{fig:shoessocks_sketch}. Any replication attempt not completed by a reset time $T_r$ is discarded and reset, irrespective of whether all actions have been performed in the correct order or not.
To what extent do resets result in canonical ordering in the conditioned ensemble of trajectories?

Concretely, we use a model in which, at each step, the system selects the canonically next action (i.e. the action remaining with smallest $i$) with probability $1 - \epsilon$ or picks uniformly among remaining actions with probability $\epsilon$. An action $j$ is in order if all actions with smaller labels $i<j$ have already been performed, and out of order otherwise. Note that the actions with smaller labels need only be completed (in any order) for action $j$ to count as in-order. In-order actions take time $\tau_\text{fast}$; out-of-order actions incur an exponentially distributed extra delay on top of $\tau_\text{fast}$, with total mean $\tau_\text{slow} = \eta_\text{stall}\,\tau_\text{fast}$ and $\eta_\text{stall} > 1$ the dimensionless stalling factor (see Fig.~\ref{fig:shoessocks_sketch}).

We measure temporal order by the fraction $q$ of action pairs performed in canonical order, rescaled as $\kappa = 2q - 1$. Thus $\kappa \in [-1,1]$: $\kappa = 1$ corresponds to canonical order, $\kappa = 0$ to a random permutation on average, and $\kappa = -1$ to the fully reversed order. We refer to $\kappa$ as the ordering score; see Supplementary Information for details. In this model, temporal ordering is the concrete form that stereotyped behavior takes. A cell that reliably performs actions in the canonical sequence has low trajectory entropy, and $\Delta S$ measures precisely how much more constrained, i.e., more stereotyped, the action sequence has become through conditioning.

\begin{figure}[h!]
  \includegraphics[width=\columnwidth]{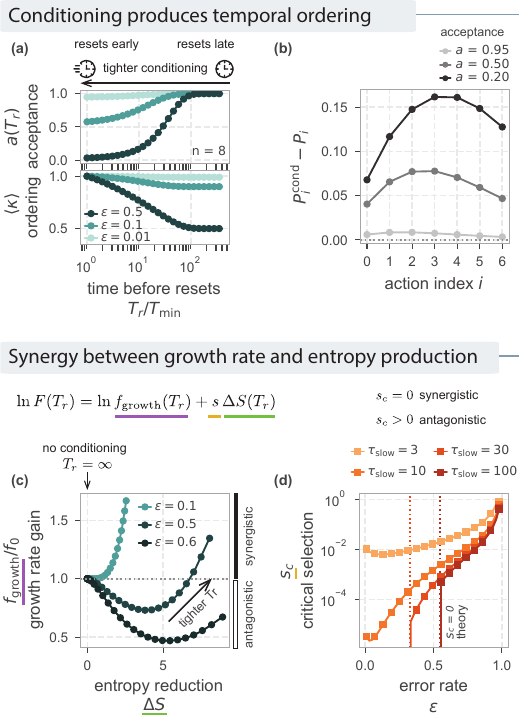}
  \caption{\textbf{Conditioning through resets produces temporal ordering in the socks-before-shoes model.} (a) Average ordering score $\langle\kappa\rangle$ vs. reset time $T_r/T_\text{min}$ (log scale) for $n = 8$, $\eta_\text{stall} = 100$ at $\epsilon = 0.01, 0.1, 0.5$. $T_\text{min}$ is the time of completion of the canonical sequence where all actions are in the correct order, $T_\text{min}=n\,\tau_\text{fast}$. Higher $\epsilon$ requires tighter resets and lower acceptance fraction $a(T_r)$ to achieve the same ordering $\kappa$. (b) The effect of conditioning on the probability that action $i$ is correct, $P_{i}$, is shown as the difference $P_{i}^\text{cond}- P_i$ vs. action index $i$ for $n = 8$, $\epsilon = 0.5$, $\eta_\text{stall} = 3$ at acceptance levels $a \approx 0.95, 0.5, 0.2$. (c) Growth-rate fold-change $f_\text{growth}/f_0$ [with $f_0$ the fitness without any conditioning, $f_0=f_\text{growth}(T_r=\infty)$] vs. $\Delta S$ for $n = 8$, $\eta_\text{stall} = 100$ at $\epsilon = 0.1$ (synergistic: curve stays above $f_\text{growth}/f_0 = 1$), $\epsilon = 0.5$ (initially antagonistic, then synergistic), and $\epsilon = 0.6$. (d) $s_c$ (critical selection for entropy reduction $\Delta S$) vs. $\epsilon$ for $\eta_\text{stall} = 3, 10, 30, 100$. At large $\eta_\text{stall}$ and small $\epsilon$, $s_c = 0$ (synergistic regime). Dotted lines mark analytically predicted boundary $\epsilon_c(\eta_\text{stall})$.}
  \label{fig:shoessocks}
\end{figure}

Fig.~\ref{fig:shoessocks}(a) shows the average ordering score $\langle\kappa\rangle$ over the ensemble of accepted trajectories conditioned on completing before $T_r$. We vary the time before the reset happens, $T_r/T_\text{min}$, where $T_\text{min}$ is the time of completion of the canonical sequence where all actions are in the correct order, $T_\text{min}=n\,\tau_\text{fast}$. Fig.~\ref{fig:shoessocks}(a) shows results for $n = 8$ actions with $\eta_\text{stall} = 100$ at error rates $\epsilon = 0.01$, $0.1$, and $0.5$. Shorter reset times (smaller $T_r$) monotonically increase ordering, with higher $\epsilon$ requiring more aggressive resets (smaller $a(T_r)$; gray lines) to achieve the same degree of order. All ordering emerges from the temporal filter alone, without the need for microscopic control over which sequence of actions is better.

Ordering is not uniform across actions. For $j>i$, let
\begin{align}
    P_{ij} = \sum_x P[x(t)] \, \mathbf{1}[i \text{ before } j \text{ in } x(t)]
\end{align}
be the probability that action $i$ is performed before action $j$, in the conditioned or unconditioned ensemble. $P_{ij}$ depends only on $i$, and write $P_{ij}\equiv P_i$. To see this, consider any two candidates $j, j' > i$. While action $i$ has not yet been performed, the dynamics do not distinguish $j$ from $j'$: the deterministic branch always picks some action $\le i$, never $j$ or $j'$, and the random branch picks each of them with equal probability. Swapping the labels $j\leftrightarrow j'$ therefore leaves the trajectory distribution unchanged, so $P_{ij}=P_{ij'}$.

Fig.~\ref{fig:shoessocks}(b) shows how this ordering probability $P_i$ changes after conditioning, $P_{i}^\text{cond} - P_{i}$, as a function of the index of the action $i$. At sufficiently tight reset thresholds, actions with middle indices benefit the most from conditioning, with early and late actions benefiting less. The first action already has a high baseline probability of being correct, $P_1 = (1-\epsilon) + \epsilon/n$, dominated by the deterministic branch $(1-\epsilon)$, which leaves little room for conditioning to improve it.

Fig.~\ref{fig:shoessocks}(c) shows the growth rate gain due to resets $f_\text{growth}/f_0$, where $f_0$ is the fitness without any conditioning $f_0=f_\text{growth}(T_r=\infty)$, versus $\Delta S$ for error rates $\epsilon = 0.1$ and $0.5$ at $\eta_\text{stall} = 100$. At low $\epsilon$, resets simultaneously increase the growth rate and impose temporal order on trajectories: the curve rises above $f_\text{growth}/f_0 = 1$ (synergistic). At high $\epsilon$, resets reduce entropy only at the cost of growth rate (antagonistic). The distinction arises because resets return the system to its initial state: at low $\epsilon$, a restarted system is likely to follow the canonical path, so trying again pays off; at high $\epsilon$, it is likely to err again, and the failed attempt is wasted.

The critical selection $s_c$ is the smallest $s$ at which the compound fitness $F(T_r)$ of Eq.~\ref{eq:3} exceeds the unconditioned $f_0$ for some $T_r$ --- i.e., the minimum selection pressure for entropy reduction at which a reset mechanism becomes favorable. Fig.~\ref{fig:shoessocks}(d) shows that $s_c$ generally decreases with lower intrinsic error rate $\epsilon$. However, for large enough $\eta_\text{stall}$, $s_c$ goes to zero at a non-zero value of $\epsilon$ indicating that a reset mechanism will evolve spontaneously and create temporal order, even without any selection for entropy reduction. This is because, in this regime, conditioning through resets improves speed in addition to ordering, a synergistic behavior. The boundary $\epsilon_c(\eta_\text{stall})$ follows from the statistical independence of each step's ordering outcome; see details in the Supplementary Information.

\begin{figure}
  \includegraphics[width=\columnwidth]{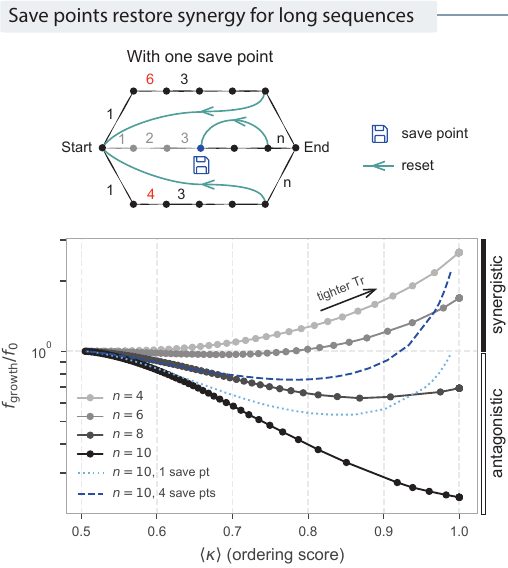}
  \caption{\textbf{System size $n$ controls the cost of conditioning, and save points restore synergy.} Growth-rate benefit from resetting $f/f_0$ vs. the ordering score $\langle\kappa\rangle$ for the permutation model of Fig.~\ref{fig:shoessocks} with $\epsilon = 0.5$, $\eta_\text{stall} = 50$. Each curve is traced out parametrically in the reset time $T_r$, with smaller $T_r$ corresponding to tighter conditioning and larger $\langle\kappa\rangle$. Gray curves show baseline (no save points) results for $n = 4, 6, 8, 10$: for small systems ($n = 4, 6$), resets simultaneously increase both temporal ordering and growth rate (synergistic regime, $f/f_0 > 1$); for large systems ($n = 8, 10$), conditioning comes at the cost of growth rate (antagonistic regime). Colored curves show the effect of save points for $n = 10$: a single midpoint save point at position $\{5\}$ (dotted light blue) partially rescues the growth rate, and save points at every second position $\{2, 4, 6, 8\}$ (dashed dark blue) restore synergistic behavior, mirroring the performance of smaller $n$ systems.}
  \label{fig:savepoints}
\end{figure}

Fig.~\ref{fig:shoessocks} looks at the effect of conditioning through resets at fixed system size, $n=8$. We now look at how the cost of resets scales with system size. Fig.~\ref{fig:savepoints} plots $f_\text{growth}/f_0$ versus $\langle\kappa\rangle$ for $n = 4, 6, 8, 10$ at $\epsilon = 0.5$, $\eta_\text{stall} = 50$ (gray curves), with each curve traced out by varying the reset time $T_r$: tighter resets (smaller $T_r$) push $\langle\kappa\rangle$ higher. For small $n$, temporal ordering and growth rate increase together (synergistic), whereas for large $n$ temporal ordering comes at the cost of growth rate (antagonistic). The transition reflects the combinatorial explosion of the permutation space: achieving a given degree of temporal ordering in a larger system requires discarding a larger fraction of trajectories, and the associated speed penalty eventually dominates.

A minimal modification recovers synergistic behavior without requiring microscopic error recognition. We introduce \emph{save points}: positions $k$ in the canonical sequence at which verified progress is preserved across resets. If a trajectory has correctly completed actions $1, \ldots, k$, future resets do not restart the entire process but rather allow resumption from action $k + 1$ rather than from scratch. The action selection process is entirely unchanged: all $n$ actions remain available at every step, and the system can still perform any action early or late. Only the reset rule differs such that the progress up to a save point is not erased. Biological assembly processes naturally exhibit such save points: phage heads and tails are assembled independently and combined \cite{yap_structure_2014}, ribosome biogenesis proceeds through metastable pre-ribosomal intermediates \cite{klinge_ribosome_2019}, and nuclear pore complexes are built from stable pre-formed subcomplexes such as the Y-complex \cite{hampoelz_structure_2019}. Save points exploit the principle that recognizing (partial) success is easier than preventing all potential failures \cite{kondev_exploratory_2026}.

The colored curves in Fig.~\ref{fig:savepoints} show the effect for $n = 10$. A single midpoint save point at position $\{5\}$ partially rescues the growth rate (light blue), whereas save points at every second position $\{2, 4, 6, 8\}$ restore fully synergistic behavior (dark blue), closely matching $n = 4$ without save points. Save points could thus allow resetting to be a viable strategy for complex multi-step processes where the time cost would otherwise be prohibitive.

\section{Save points in templated replication}
\begin{figure*}[t]
  \includegraphics[width=\textwidth]{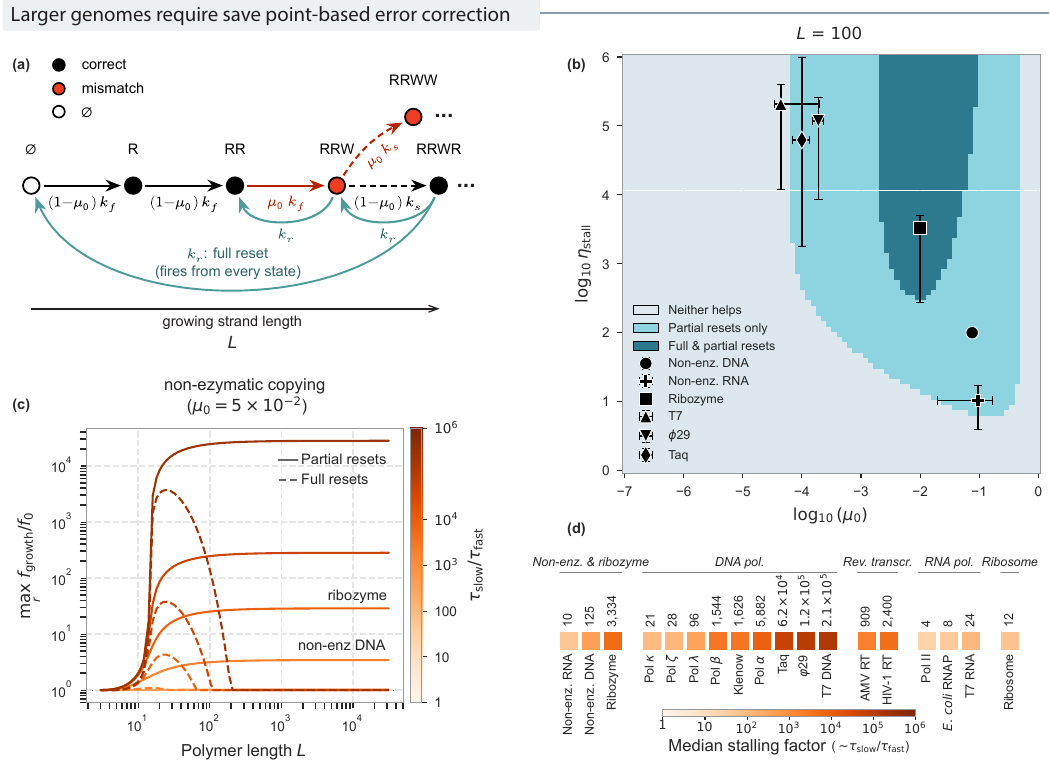}
  \caption{\textbf{Exonuclease proofreading implements save points and is necessary to copy larger genomes.} \textbf{(a)}~Sketch of the polymerization model: monomers are added at rate $k_\text{fast}$, with incorrect incorporations (probability $\mu_0$) triggering stalling (rate reduced by $\eta_\text{stall}$). Partial resets (exonuclease) cleave the last monomer; full resets return to $n=0$. \textbf{(b)}~Categorical heatmap of the $(\mu_0,\,\eta_\text{stall})$ plane for $L=100$, classifying each parameter combination by whether partial resets, full resets, or neither improves the growth rate beyond the uncorrected baseline. Markers show measured values for non-enzymatic DNA and RNA copying, ribozymes, and DNA polymerases (T7, Taq, and $\phi29$ exo$^-$ variants). \textbf{(c)}~Maximum growth-rate improvement $\max_r f_\text{growth}/f_0$ vs.\ polymer length $L$ at $\mu_0=0.05$ (non-enzymatic regime). Solid lines: partial resets; dashed lines: full resets. Partial resets plateau at large $L$, while full resets peak then decay---discarding $O(L)$ progress becomes increasingly costly for longer polymers. Color encodes $\eta_\text{stall}$. \textbf{(d)}~Median experimental stalling ratios for representative replication systems, ordered by complexity; see \cite{ravasio_evolution_2026} for a comprehensive list of references and \cite{jia_diaminopurine_2024} for non-enzymatic RNA copying. The per-nucleotide error rates $\mu_0$ overlaid as markers in (b) are listed in Supplementary Table~\ref{tab:mu_sources} together with their references.} 
  \label{fig:poly}
\end{figure*}
Resets with save points provide a natural framework for analyzing real biological error-correction in the context of, e.g., the templated replication of nucleic acids. DNA polymerases often contain an exonuclease domain that cleaves off nucleotides from the tip of the growing strand, while RNA polymerases are known to backtrack and excise a stretch of recently added nucleotides. Each of these mechanisms are effectively a partial reset \cite{roldan_stochastic_2016} that requires the polymerase to re-synthesize the cleaved nucleotide.

We implement a stochastic model of templated replication with both partial resets (exonuclease-like resets with save points) and full resets; see Fig.~\ref{fig:poly}(a). The strand is elongated either by adding a correct nucleotide with rate $(1-\mu_0)k_\text{fast}$ or by adding a mismatched nucleotide with rate $\mu_0\,k_\text{fast}$. We set the timescales by fixing $k_\text{fast}=1$. The elongation after a mismatch is stalled with a rate $k_\text{slow}$ and partial, or full, resets happen with a rate $k_{r}$.

The model parameters are biologically interpretable. The probability $\mu_0$ that the wrong nucleotide is incorporated is set by the physical chemistry of nucleotide recognition. For non-enzymatic copying this is due entirely to the hydrogen bonds formed in base-pairing. The free energy difference between a matched and mismatched base-pair is estimated to be $\Delta G \approx 1$ to $3 \, k_B T$, giving $\mu_0 = e^{-\Delta G} \approx 10^{-2}$. For enzymatic copying by a polymerase this is more complex due to nucleotide selectivity at the polymerase active site. However, we may estimate it from measured values of the mutation rate of DNA polymerases lacking an exonuclease domain, giving $\mu_0 \sim 10^{-5}$--$10^{-4}$, with exo-deficient $\phi29$ variants falling near the upper end of this range.

Similarly, the slow down after a mismatch, $\eta_\text{stall} \equiv k_\text{fast}/k_\text{slow}$, arises from the widely observed `stalling' of replication after the incorporation of a mismatch. The stalling factors $\eta_\text{stall}$ have been measured for several processes across the central dogma, Fig.~\ref{fig:poly}(d). These range from $\eta_\text{stall} \approx 10^1$ for non-enzymatic replication of RNA, to $\eta_\text{stall} \approx 10^5$ for high-fidelity, replicative DNA polymerases such as that of T7.

In the context of templated replication, `partial resets' require the action of an exonuclease domain or some similarly complex machinery. We sought to understand when evolution would favor such machinery if the only selection were for faster replication. We calculated the change to the growth rate, $f_\text{growth}/f_0$, over a range of error rates $\mu_0$ and stalling factors $\eta_\text{stall}$ for a small strand of length $L = 100$, relevant to origins-of-life scenarios, with either full or partial resets. On this data, Fig.~\ref{fig:poly}(b), we overlaid the experimentally measured values of stalling factors and bare error rates for non-enzymatic copying of DNA and RNA, a ribozyme, and three DNA polymerase systems (T7, Taq, and exo-deficient $\phi29$ variants). We find that, even for a strand as small as $100$ nucleotides, the more complex partial resets are favored over the simpler-to-implement full resets over a broad range of experimentally relevant parameters.

We then varied the length of the strand $L$, plotting in Fig.~\ref{fig:poly}(c) the best gain in speed ($\max_{k_r} f_\text{growth}/f_0$) for both partial and full resets with $\mu_0 = 5\times10^{-2}$ (relevant to, e.g., non-enzymatic copying at the origins of life). We find that partial resets are beneficial for any sufficiently large genome sizes $L$. In contrast, the mechanistically simpler full resets are only beneficial in a narrow range of genome sizes between $L \sim 10^1$ and $L \sim 10^2$ nucleotides. These results together suggest that partial resets, as implemented by the complex machinery of an exonuclease, are necessary for conditioning to evolve `for free' when selecting for the replication speed to copy a large genome.

\section{Phenomenology of $P(T)$}
The socks-before-shoes model showed that whether conditioning is synergistic or antagonistic hinges on one parameter: how much slower disordered trajectories are relative to ordered ones ($\eta_\text{stall}$). When $\eta_\text{stall}$ is large, the fastest-growing population automatically becomes the most ordered ($s_c = 0$); when $\eta_\text{stall}$ is small, order comes only at a growth-rate cost ($s_c > 0$). In most biological systems, $\eta_\text{stall}$ is not directly observable; what can be characterized is the completion-time distribution $P(T)$ of the process as a whole. A heavy-tailed $P(T)$, in which a small fraction of attempts take dramatically longer than typical, plays the same role as large $\eta_\text{stall}$: resetting those slow outliers is inexpensive and effective. We now ask how the relationship between growth rate $f_\text{growth}$ and entropy reduction $\Delta S$ depends on the shape of $P(T)$ across five standard distribution families.

We first use the log-normal family for $P(T)$ to illustrate the key phenomenology. All distributions are normalized to unit mean so that differences reflect shape alone. Definitions, parameter ranges, and full numerical details for five standard families are given in the Supplementary Information, Fig.~\ref{fig:SI_concept_families}.

\begin{figure}[t]
  \includegraphics[width=\columnwidth]{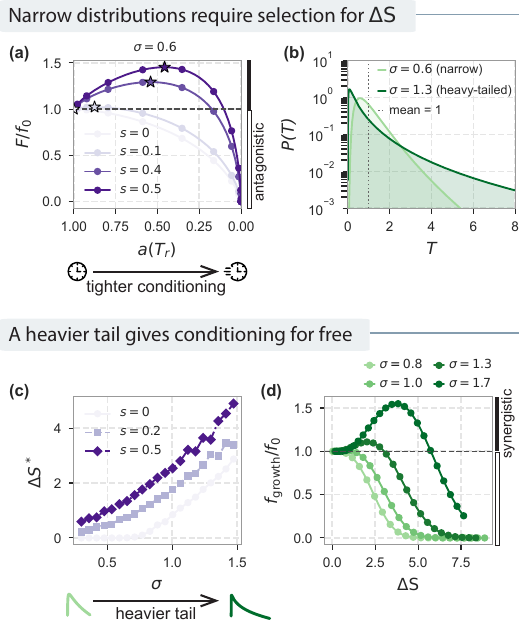}
  \caption{\textbf{Distribution shape controls the cost and benefit of resets.} All distributions are lognormal, normalized to unit mean. (a) Compound fitness $F/f_0$ vs. acceptance fraction $a(T_r)$ for $\sigma = 0.6$ (narrow shape) at several selection strengths $s = 0, 0.1, 0.4, 0.5$. Large dots mark the optimal acceptance fraction $a^*$ for each $s$. At $s = 0$ (selection for speed only), resets always cost; increasing $s$ shifts the optimum toward tighter resets (smaller $a$). (b) Lognormal completion-time distributions on a log scale for narrow ($\sigma = 0.6$) and heavy-tailed ($\sigma = 1.3$) cases, illustrating how tail weight sets the potential for resets to improve speed. (c) Optimal entropy reduction $\Delta S^*$ vs. shape parameter $\sigma$ for selection strength $s = 0, 0.2, 0.5$. Heavier tails (larger $\sigma$) enable spontaneous stereotyping ($\Delta S^* > 0$) even at $s = 0$; lighter tails require non-zero selection pressure $s$ for entropy reduction. (d) Speed-vs-entropy trade-off: $f_\text{growth}/f_0$ (growth-rate fold-change) vs. $\Delta S$ (entropy reduction), parametric in $T_r$, for $\sigma = 0.5$ to $1.8$. Narrow distributions (small $\sigma$) show an antagonistic regime where stereotyping reduces speed; heavy-tailed distributions (large $\sigma$) show a synergistic regime where stereotyping and speed increase together.}
  \label{fig:lognormal}
\end{figure}

Fig.~\ref{fig:lognormal}(a) shows the compound fitness $F$ as a function of acceptance fraction $a(T_r)$ for a log-normal ($\sigma = 0.6$) $P(T)$ for several selection strengths $s = 0, 0.1, 0.4, 0.5$ for entropy reduction. At $s = 0$ (pure growth rate selection), the fitness is maximized when all trajectories are accepted: resets reduce growth rate and are never favored. As $s$ increases, the stereotyping benefit offsets the reduction in growth rate; the optimal acceptance fraction shifts toward stronger resets (smaller $a$), with large dots marking the fitness peak for each $s$.

These results depend on the shape of $P(T)$. Varying $\sigma$ in the log-normal family changes the tail weight [Fig.~\ref{fig:lognormal}(b)], which in turn changes how the acceptance fraction $a(T_r)$ affects both growth rate and entropy reduction. For heavy-tailed distributions, discarding a fraction of slow completions has a qualitatively different cost-benefit balance than for light-tailed ones, altering which reset strategies are favored.

To quantify how much stereotyping resets can produce, we find the reset time $T_r^*$ that maximizes the compound fitness $F$ [as in Fig.~\ref{fig:lognormal}(a)], then determine the entropy reduction $\Delta S^* = \Delta S(T_r^*)$ achieved at that optimum. Fig.~\ref{fig:lognormal}(c) traces the optimal entropy reduction $\Delta S^*$ as a function of the shape parameter $\sigma$ for $s = 0, 0.2, 0.5$. For non-zero $s$, resets are favored across a wide range of $\sigma$: the heavier the tail, the less selection pressure is needed to justify them. At $s = 0$, a critical lognormal shape $\sigma_c \approx 0.8$ emerges: above it, $\Delta S^* > 0$ without any selection for entropy reduction at all. Beyond this threshold, the growth rate benefit of resetting heavy-tailed distributions is large enough that stereotyped behavior emerges spontaneously as a byproduct. The trajectory ensemble becomes more reproducible not because reproducibility is selected for, but because the fastest growth strategy happens to discard many paths through conditioning.

Fig.~\ref{fig:lognormal}(d) highlights these two behaviors. Each curve traces the parametric path $(\Delta S,\; f_\text{growth}/f_0)$ as $T_r$ varies, for four values of $\sigma$. Heavy-tailed distributions (large $\sigma$) produce curves that rise above $f_\text{growth}/f_0 = 1$ while $\Delta S$ increases: a synergistic regime where resets simultaneously speed growth and impose stereotyped behavior. Light-tailed distributions (small $\sigma$) produce curves that dip below 1 as $\Delta S$ grows: an antagonistic regime where stereotyping comes at a growth rate cost. The boundary between these regimes corresponds to $\sigma_c$.

\begin{figure}[b]
  \includegraphics[width=\columnwidth]{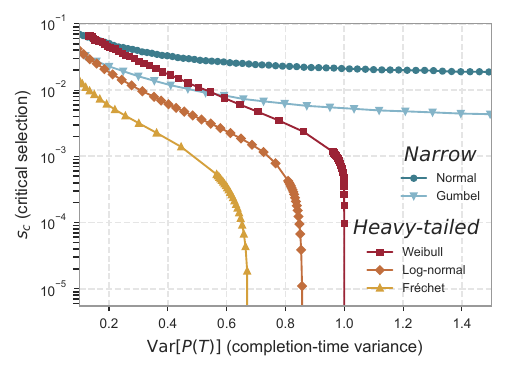}
  \caption{\textbf{The shape of the tail of $P(T)$ determines whether resets require selection for entropy reduction.} Critical selection for entropy reduction $s_c$  vs. completion-time variance $\mathrm{Var}[P(T)]$ for five distribution families, all normalized to unit mean; for each family the shape parameter is varied to sweep variance. Three families (Fr\'{e}chet, Log-normal, Weibull) exhibit a sharp transition to $s_c = 0$ at a family-specific critical variance, indicating that resets become spontaneously favorable as a byproduct of selection for growth rate alone. The transition occurs at progressively higher variance for lighter-tailed families: Fr\'{e}chet (power-law tail) transitions first, followed by Log-normal, then Weibull (which crosses from light- to heavy-tailed as its shape parameter $k$ decreases through 1). Normal and Gumbel require positive $s_c$ at all variances, though both decrease with increasing variance. The Fr\'{e}chet curve covers only the finite-variance regime ($\alpha > 2$); for smaller $\alpha$ (including the divergent-variance regime), $s_c$ remains zero, see Supplementary Information.}
  \label{fig:survey}
\end{figure}

The same analysis extends beyond the log-normal. Fig.~\ref{fig:survey} shows the critical selection $s_c$ as a function of completion-time variance for five standard families (Normal, Weibull, Log-normal, Fr\'{e}chet, Gumbel), all normalized to unit mean. Heavy-tailed families (Fr\'{e}chet; Log-normal above $\sigma_c$; Weibull at low shape parameter $k$) exhibit a sharp transition to $s_c = 0$ at a family-specific critical variance, indicating that resets become spontaneously favorable from growth rate benefits alone. The transition occurs at progressively higher variance for lighter-tailed families. Normal and Gumbel distributions, whose tails decay exponentially or faster, require positive $s_c$ at all variances, though the required pressure decreases as variance grows.

Fig.~\ref{fig:survey} shows three regimes: spontaneous stereotyping ($s_c = 0$, heavy tails), pressure-dependent stereotyping ($s_c > 0$, light tails), and universal stereotyping (large $s$, any distribution). The critical pressure $s_c$ depends on distribution shape alone (at fixed mean) and answers a concrete physical question: how heavy must the tails be for the growth rate benefit of resets to pay for stereotyping on its own?

While the survey in Fig.~\ref{fig:survey} characterizes the phenomenology across distribution families, the framework also makes precise quantitative predictions for systems with specific molecular mechanisms. A concrete example is templated replication with stochastic errors: each error-free nucleotide incorporation takes time $\tau_\text{fast}$, but errors induce stalling parameterized by the dimensionless stalling factor $\eta_\text{stall} = \tau_\text{slow}/\tau_\text{fast}$. For a sequence of length $L$ with per-nucleotide error rate $\mu_0$, the completion time distribution is well approximated by a mixture of two delta functions: a fraction $(1-\mu_0)^L$ of fast trajectories at time $L\,\tau_\text{fast}$, and the remaining fraction $1-(1-\mu_0)^L$ of slow trajectories at time $\approx(L-1+\eta_\text{stall})\,\tau_\text{fast}$. This is a coarse-grained version of the polymerization model of Fig.~\ref{fig:poly}. Supplementary Fig.~\ref{fig:SI_twodelta} shows how the fold-change in growth rate with optimal resets depends on the error rate $\mu_0$ and the stalling factor $\eta_\text{stall}$.

\section{Discussion}
Conditioning is a counter-intuitive route to stereotyping trajectories in stochastic high-dimensional systems. Rather than controlling a system at small scales, a mechanism removes any physical record of trajectories that fail to meet a criterion, here, the completion by a threshold time $T_r$. In the spirit of Jaynes \cite{jaynes_information_1957}, conditioning on a final outcome makes the surviving trajectories appear purposeful \cite{chetrite_nonequilibrium_2015}, as though effective forces prevent the system from taking a wrong turn, even though the microscopic dynamics are unchanged. In the socks-before-shoes model, conditioned trajectories reliably follow the canonical sequence not because any mechanism directs them to, but because the temporal filter has pruned those that don't.

We have quantified the minimum selection pressure $s_c$ for reducing trajectory entropy $\Delta S$ needed to evolve conditioning through resets as a stereotyping mechanism. We find that when the variance of completion times $T$ is large, even modest selection suffices to evolve conditioning. In fact, when the distribution $P(T)$ is wide enough, there is no selection needed for reducing the entropy of trajectories ($s_c = 0$) and conditioning evolves spontaneously as a byproduct of fast exponential replication, with no selection for entropy reduction required.

The conditioning mechanism we study is mechanistically simple: it requires no recognition of specific errors, only a temporal signal. As suggested earlier \cite{kondev_exploratory_2026,Winfree2019-uj}, recognizing and stabilizing successful outcomes, including partial success in the case of save points, is mechanistically simpler than specifying the correct trajectory. Moreover, in many biochemical processes, temporal signals naturally correlate with trajectory quality: trajectories that deviate from functional paths tend to proceed more slowly, e.g., through stalling \cite{ravasio_evolution_2026}. This correlation appears across systems spanning the central dogma. A mismatched nucleotide slows polymerase extension \cite{perrino_differential_1989,ravasio_evolution_2026}, a misfolded intermediate gets trapped before completing folding, and a defective subunit holds up further assembly \cite{evans_physical_2017,evans_optimizing_2018, winfree_proofreading_2004,jhaveri_discovering_2024}. In such cases, conditioning through resets not only reduces trajectory entropy but also biases the observable ensemble toward biochemically functional or desirable paths.

The physical mechanism of conditioning operates through destruction of failed trajectory products. Products, such as polymers, structures, or folded proteins, are physical records of their assembly trajectories, so destroying incomplete or slow outcomes erases the record of disordered paths, and only the conditioned ensemble persists (see Supplementary Information Section~\ref{sec:SI-physical}). The physical requirements are minimal: displacement from a substrate by competing molecules \cite{Mukherjee2024-ij}, spatial confinement \cite{phillips_molecular_2020}, temperature cycling, or simple degradation of intermediates could all implement conditioning \cite{matsubara_kinetic_2018,matsubara_avoidance_2023}. In the context of the origin of life, such temporal conditioning may have been among the earliest mechanisms accessible to primitive replicators \cite{Chen2004-oh,tjhung_rna_2020, obermayer_emergence_2011,goppel_kinetic_2021}, requiring only a stochastic completion process subject to some form of resetting, long before complex molecular recognition could evolve. Our framework quantifies when this broadly accessible strategy is favorable in terms of the shape parameters of the distribution of completion times $P(T)$.

Even if the temporal signal did not correlate with trajectory quality in this detailed biochemical sense, conditioning would still provide value through entropy reduction alone. Stereotyped assembly trajectories are a prerequisite for downstream functionalization, as in glycan branching patterns where reproducible structures enable specific molecular recognition \cite{lau_complex_2007,dennis_metabolism_2009}, or in protein complexes where consistent stoichiometry and geometry determine function \cite{monod_nature_1965,doyle_structure_1998,lander_complete_2012}. Only relatively reproducible phenotypes are selectable in Darwinian evolution \cite{siegal_waddington_2002} since high trajectory entropy makes outcomes too variable for selection to act on consistently. Our framework rewards the reduction of trajectory entropy, $\Delta S$, rather than adherence to a particular ``correct'' path, and so applies when stereotypy itself is the functional output.

Stochastic resets have been extensively studied in the context of search and first-passage problems \cite{evans_diffusion_2011,evans_stochastic_2020,reuveni_role_2014,reuveni_optimal_2016,pal_first_2017} and in the context of exploratory dynamics \cite{kondev_exploratory_2026}. Our work builds on this body of work in several respects. First, we focus on the reduction of trajectory entropy, not just the speed-up of completion. Prior work on entropy in the resets context \cite{fuchs_stochastic_2016,busiello_entropy_2020,eliazar_entropy_2023,louwerse_information_2022} considered the entropy of the completion time distribution or the thermodynamic cost of resetting; our interest is in the entropy of paths, which we connect to observable stereotypy in biological products. We consider an overall fitness that combines speed and entropy reduction, and study their relative contributions to determine how strongly entropy reduction must be weighted for conditioning to evolve. We also situate resets in the context of self-replication, where fitness is defined by the Euler-Lotka equation rather than a simple mean first-passage time. Finally, we identify the hierarchical structure of temporal ordering produced by such resets in a mechanistic model (socks-before-shoes) and propose mechanistically inexpensive modifications (save points) that can maintain conditioning benefits at lower cost.

Our results provide a non-equilibrium counterpart to Waddington's canalization, the developmental channeling of trajectories into reproducible paths \cite{waddington_canalization_1942,waddington_strategy_1957}. In our work, canalization is achieved not through an energy landscape that shapes the entire trajectory, but through a non-equilibrium mechanism (resets) that prunes disordered trajectories from the observable ensemble. Just as Waddington's canalization produces phenotypic robustness by constraining developmental paths, conditioning through resets produces trajectory stereotypy by constraining the observable ensemble. The mechanism differs, but the outcome of reproducible trajectories despite underlying stochasticity is the same.

Our work leaves several questions unanswered. The interplay between reset-based conditioning and direct error recognition remains to be understood quantitatively, e.g., proofreading in polymerases is thought to be a mix of simple conditioning and direct error recognition working in concert \cite{midha_synergy_2023,midha_insights_2024}. The framework also generalizes beyond temporal conditioning: any observable correlated with trajectory quality (size, shape, composition) could serve as the basis for conditioning, not just completion time. Similarly, other models of partial resets, besides save points studied here, could reduce the time cost of conditioning while preserving the ordering benefit. Finally, connecting this framework to specific experimental systems (e.g., in vitro assembly of macromolecular complexes \cite{perlmutter_mechanisms_2015,winfree_proofreading_2004}, RNA and DNA replication in the presence of strands competing for the template, co-transcriptional ribosomal assembly where checkpoints gate subunit maturation \cite{sanghai_co-transcriptional_2023}), where both the distribution of completion times and the stereotypy of products can be measured, would provide a direct test of our predictions.

\begin{acknowledgments}
We thank J. Kondev, E. Winfree, M. Kirschner and members of the Murugan group and the CZI theory group for discussions. We are grateful to Julie Theriot for her inspiring thoughts about socks-before-shoes kinetics. This research was supported by National Science Foundation through the Center for Living Systems (grant no. 2317138) and through grants from the NSF (DMS-2235451) and Simons Foundation (MPS-NITMB-00005320) to the NSF-Simons National Institute for Theory and Mathematics in Biology (NITMB). R.R. was supported by the NIGMS of the National Institutes of Health under award number R35GM151211 and the NSF (PHY-2310781). R.P. acknowledges NIH Maximizing Investigators’ Research Award (MIRA) 1R35 GM118043-01. A.M. and J.W.S. received support from the Sloan (G-2022-19518) and Moore (11479) foundations Matter-to-Life program.
\end{acknowledgments}

\section*{Data and code availability}
All code used to generate the figures and numerical results in this work is available at \url{https://github.com/riccardoravasio/just-a-clock} \cite{ravasio_just_a_clock_code}.

\putbib
\end{bibunit}

\clearpage
\onecolumngrid

\begin{center}
{\Large\bfseries Supplementary Information: \\[0.3em] Conditioning as a route to stereotyped behavior in growing populations}
\end{center}
\vspace{1em}

\begin{bibunit}

\setcounter{figure}{0}
\setcounter{table}{0}
\setcounter{equation}{0}
\setcounter{section}{0}
\renewcommand{\thefigure}{S\arabic{figure}}
\renewcommand{\thetable}{S\arabic{table}}
\renewcommand{\theequation}{S\arabic{equation}}
\renewcommand{\thesection}{S\arabic{section}}
\renewcommand{\thesubsection}{\thesection.\arabic{subsection}}
\renewcommand{\thesubsubsection}{\thesubsection.\arabic{subsubsection}}
\makeatletter
\renewcommand{\p@subsection}{}
\renewcommand{\p@subsubsection}{}
\makeatother
\setcounter{secnumdepth}{3}

\section{General framework}
\label{sec:SI-general}

This section develops the mathematical framework that underlies the rest of the paper. The central objects are a population of self-replicating systems, the distribution of times they take to complete replication, and the effect of discarding (resetting) attempts that take too long. We derive how resets change both the population growth rate and the reproducibility of the replication process, and define a combined fitness that accounts for both.

\subsection{Setup: replication through stochastic trajectories}

We consider a self-replicating system that must traverse a stochastic trajectory $x(t)$ from an initial state (Birth) to a final state (Division). The state space is high-dimensional. Depending on the system of interest, $x(t)$ might track the assembly of a macromolecular structure, the copying of genetic information, or other biochemical tasks needed for division. Each path $x(t)$ is realized with probability $P[x(t)]$, and the completion time $T[x] = T$ is a functional of the trajectory. The completion-time distribution $P(T)$ is the marginal of $P[x(t)]$ induced by the map $x \mapsto T[x]$. Broad completion-time distributions arise naturally from kinetic traps and frustrated states \cite{lin_effects_2017,lahini_nonmonotonic_2017,lenz_geometrical_2017}.

\subsection{Fitness as long-term exponential growth rate}

Consider a self-replicating system with replication-time distribution $P(T)$. The fitness $f_\text{growth}$ is the long-term exponential growth rate,

\begin{equation}
\langle N(t) \rangle \sim e^{f_\text{growth}\,t},
\label{eq:exp-growth}
\end{equation}

where $N(t)$ is the population size after time $t$. The fitness is \emph{not} simply $1/\langle T \rangle$. In a growing population, a single slow replicator does not hold back the rest since its sister cells continue dividing independently. As a result, fitness is dominated by fast replicators, not penalized by slow ones.

To derive $f_\text{growth}$, let $N(t)$ be the population of a lineage seeded by one cell, with generation times drawn independently from $P(T)$. The expected population satisfies the renewal equation

\begin{equation}
\langle N(t) \rangle = 2 \int_0^\infty dT\, P(T)\, \langle N(t - T) \rangle,
\label{eq:renewal}
\end{equation}

where the factor of 2 accounts for the two daughter cells produced at each division. The idea is simple: the founding cell divides at some time $T$ drawn from $P(T)$, producing two daughters, each of which then seeds its own growing subpopulation of expected size $\langle N(t-T)\rangle$. Substituting the exponential growth ansatz $\langle N(t) \rangle \sim e^{f_\text{growth}\,t}$ yields the Euler-Lotka equation \cite{sharpe_problem_1911,jafarpour_bridging_2018}

\begin{equation}
2\,\tilde{P}(-f_\text{growth}) = 1,
\label{eq:euler-lotka}
\end{equation}

where $\tilde{P}(w) = \int_0^\infty e^{wT} P(T)\, dT$ is the moment generating function (MGF) of $P(T)$. Evaluated at $w = -f_\text{growth}$, this is a Laplace transform: $\tilde{P}(-f_\text{growth}) = \int_0^\infty e^{-f_\text{growth}\,T} P(T)\, dT$. The exponential factor $e^{-f_\text{growth}\,T}$ downweights slow replicators and upweights fast ones, capturing the intuition that in a growing population, early divisions contribute disproportionately to long-term growth.

\subsection{Conditioning reduces trajectory entropy}

In many biological processes, incomplete or defective products are destroyed, degraded, or outcompeted, so only trajectories completing within some window are observed. We formalize this as conditioning on completion before a time $T_r$.

\subsubsection{Trajectory entropy reduction} The fundamental measure of stereotyping is the reduction in trajectory entropy. The full trajectory entropy is

\begin{equation}
H[P] = -\int P[x(t)] \ln P[x(t)]\, \mathcal{D}x(t),
\label{eq:traj-entropy}
\end{equation}

and the conditioned trajectory distribution $P^\text{cond}[x(t)] = P[x(t) \mid T[x] < T_r]$ \cite{touchette_large_2009,chetrite_nonequilibrium_2015} has entropy $H[P^\text{cond}]$. The trajectory entropy reduction

\begin{equation}
\Delta S(T_r) \equiv H[P] - H[P^\text{cond}]
\label{eq:delta-S}
\end{equation}

quantifies the stereotyping imposed by conditioning. Support restriction by itself does not prove $\Delta S \geq 0$ for an arbitrary distribution, because the surviving probabilities are renormalized. In the models and parameter ranges used in this work, the temporal filter preferentially removes slow, disordered trajectories, so the computed Shannon difference is non-negative and $\Delta S$ is a genuine entropy reduction.

\subsubsection{Completion-time entropy as a distribution-level proxy} Let $a(T_r) = \int_0^{T_r} P(T)\, dT$ denote the acceptance fraction, the probability that a randomly chosen trajectory finishes before the deadline $T_r$. The conditioned completion-time distribution is

\begin{equation}
P^\text{cond}(T) = \frac{P(T)}{a(T_r)} \quad \text{for } T < T_r, \qquad 0 \text{ otherwise.}
\label{eq:P-cond}
\end{equation}

The completion-time entropy difference $H_T[P] - H_T[P^\text{cond}]$, where $H_T[P] = -\int P(T) \ln P(T)\, dT$, is the distribution-level ordering measure used when only $P(T)$ is specified. In the socks-before-shoes model (Section~\ref{sec:SI-shoessocks}), trajectories are discrete permutations and we compute the trajectory entropy difference $\Delta S$ by summing over permutations. In the distribution family survey (Section~\ref{sec:SI-distributions}), only the completion-time distribution $P(T)$ is specified, so $\Delta S$ denotes this completion-time Shannon entropy difference. This is a proxy for trajectory-level stereotyping rather than a general lower bound on it. In the two-delta model (Section~\ref{sec:SI-applications}), the completion-time and path-type descriptions coincide because each path type has a deterministic completion time.

\subsubsection{Relation to prior work on restart entropy} Eliazar and Reuveni \cite{eliazar_entropy_2023} established universal criteria for when sharp restart increases or decreases the Shannon entropy of the completion-time distribution. Our focus is complementary: we ask whether conditioning produces ordered, reproducible trajectories in a high-dimensional state space, and embed this question in the context of self-replicating populations governed by the Euler-Lotka equation. The present work connects the mathematical foundations of restart entropy to observable physical order and to the evolutionary dynamics of growing populations.

\subsection{Resets as a physical implementation of conditioning}

A reset mechanism implements conditioning by destroying an incomplete product at time $T_r$ and restarting the process. Unlike a classical Maxwell demon, this requires only a clock, not knowledge of the microscopic state.

\subsubsection{Conditioned distribution} With resets at $T_r$, the distribution of completion times conditioned on completion before $T_r$ is

\begin{equation}
P^r(T) = \frac{P(T)}{a(T_r)} \quad \text{for } 0 \leq T < T_r, \qquad 0 \text{ otherwise,}
\label{eq:P-reset-cond}
\end{equation}

with MGF (moment generating function)

\begin{equation}
\tilde{P}^r(w) = \frac{1}{a(T_r)} \int_0^{T_r} e^{wT} P(T)\, dT.
\label{eq:MGF-cond}
\end{equation}

\subsubsection{Distribution of total replication times} A replication event consists of some number of failed attempts (each lasting $T_r$) followed by one successful completion. Each attempt independently succeeds with probability $a$ or fails with probability $1-a$, so the number of failed attempts before success follows a geometric distribution. The probability of exactly $n$ resets before success is $(1-a)^n \cdot a$, giving

\begin{equation}
P^\text{reset}(T) = \sum_{n=0}^{\infty} (1-a)^n\, a\, P^r(T - nT_r).
\label{eq:P-reset-total}
\end{equation}

The MGF follows from summing the geometric series:

\begin{equation}
\tilde{P}^\text{reset}(w) = \frac{a_w(T_r)}{1 - (1-a)\, e^{wT_r}},
\label{eq:MGF-reset}
\end{equation}

where $a_w(T_r) = \int_0^{T_r} e^{wT} P(T)\, dT$ is a weighted acceptance fraction (reducing to $a$ when $w = 0$). This decomposition into geometric restart cycles follows the renewal approach developed in the stochastic restart literature \cite{pal_first_2017}.

\subsubsection{Growth rate with resets} Substituting into the Euler-Lotka equation $2\tilde{P}(-f_\text{growth}) = 1$ yields

\begin{equation}
2 \int_0^{T_r} e^{-f_\text{growth}\,T} P(T)\, dT = 1 - (1 - a(T_r))\, e^{-f_\text{growth}\,T_r}.
\label{eq:euler-lotka-reset}
\end{equation}

This implicit equation determines the growth rate $f_\text{growth}(T_r)$ for any $P(T)$ and reset time $T_r$. The left-hand side counts the Laplace-weighted contribution of successful attempts; the right-hand side accounts for the time lost to failed attempts through the factor $e^{-f_\text{growth}\,T_r}$.

\subsection{Combined fitness: growth rate and trajectory entropy reduction}

Conditioning via resets has two effects: it modifies the growth rate $f_\text{growth}(T_r)$, and it reduces the trajectory entropy by $\Delta S(T_r)$, producing more reproducible outcomes. If reproducibility has fitness value, these combine into an effective fitness

\begin{equation}
\ln F = \ln f_\text{growth} + s\, \Delta S,
\label{eq:combined-fitness}
\end{equation}

where $s \in [0, 1)$ is the selection coefficient for trajectory entropy reduction, and $\Delta S = H[P] - H[P^\text{cond}]$ is the Shannon entropy reduction. At $s = 0$, fitness reduces to pure growth-rate selection; as $s$ increases, reproducibility (through trajectory entropy reduction) becomes increasingly important.

\subsubsection{Why this form is multiplicative, not additive} One might consider an additive alternative, $F_\text{add} = f_\text{growth} + s\, \Delta S$, treating growth rate and order as interchangeable goods. This form breaks down in the continuous tight-conditioning regime: as $T_r \to 0$, the Shannon entropy reduction $\Delta S$ can diverge because conditioning selects an ever-narrower slice of trajectories, while $f_\text{growth} \to 0$ because nothing finishes in time. An additive fitness would diverge for any $s > 0$, implying that the optimal strategy is to accept almost nothing. The multiplicative (log-additive) form avoids this because the $f_\text{growth}$ prefactor ensures that order can only enhance growth, not replace it: an organism that never replicates has zero fitness regardless of how ordered its (nonexistent) products would be.

\subsubsection{Interpretation of $s$ and the scaling at small $T_r$}
\label{sec:SI-s-range}
The parameter $s$ quantifies the per-generation fitness benefit of entropy reduction: each additional nat of Shannon entropy reduction multiplies the effective fitness by $e^s$. Throughout the calculations, $\Delta S$ is computed from the definition $\Delta S = H[P] - H[P^\text{cond}]$. Suppose the accepted set has probability $a(T_r)$ and, as $T_r \to 0$, the density on the accepted set is regular enough that the entropy of the normalized local shape remains bounded. Then conditioning amplifies the accepted density by $1/a$, giving $H[P^\text{cond}] = H[P|_{T<T_r}] + \ln a$ with a bounded local-shape term, and therefore the Shannon entropy difference obeys $\Delta S = -\ln a + O(1)$. This is exact for uniform conditioning on a fraction $a$ of the support and is only an asymptotic statement otherwise.

Under the same small-acceptance limit, expanding the Euler-Lotka equation gives $f_\text{growth} \sim a/T_r$: the mean replication time is roughly $T_r/a$, set by the $\sim 1/a$ attempts needed before one succeeds. Combining this growth-rate scaling with the continuous entropy asymptotic gives $F = f_\text{growth}\, e^{s\Delta S} \sim (a/T_r)\, a^{-s}$ up to bounded prefactors, so the combined fitness scales as

\begin{equation}
F(T_r) \sim \frac{a(T_r)^{1-s}}{T_r}.
\label{eq:F-scaling}
\end{equation}

For $s < 1$, this asymptotic suppresses arbitrarily tight conditioning: a non-replicating organism cannot achieve high fitness merely by selecting an extremely rare ordered subset. For $s \geq 1$, the asymptotic entropy reward could dominate the vanishing growth rate in this idealized limit, which is biologically meaningless. Thus $s < 1$ is a consistency requirement for the continuous tight-conditioning regime, while all numerical results use the directly computed Shannon entropy reduction.

\subsubsection{When does reproducibility have intrinsic fitness value?} The ordering term in $F$ rewards low trajectory entropy, not adherence to a particular ``correct'' path. Low entropy means the system reliably produces the same outcome, and such reliability can be functionalized by downstream processes precisely because it is predictable. In many biological systems, errors correlate with slowness (stalling, kinetic traps), so the reliable ensemble selected by temporal conditioning coincides with what would conventionally be called ``correct.'' But our framework does not require this identification and hence can also be applied to contexts where stereotypy is a precondition for function rather than a consequence of it.

Cell-surface glycans serve as molecular identity markers; the particular sugar pattern is evolutionarily arbitrary, but it must be reproducible so the immune system can distinguish self from non-self \cite{varki_biological_2017}. Foraging ants converge on a single trail because a consistently reinforced trail is followable, while a diffuse cloud of pheromone provides no directional information \cite{camazine_self-organization_2001}. A mirror-image biochemistry built on D-amino acids would be equally functional, but a mixture of L and D prevents stable protein folding \cite{blackmond_origin_2010}. Multi-subunit self-assembly (viral capsids, microtubules) requires uniform subunit conformations; variable subunits produce amorphous aggregates \cite{perlmutter_mechanisms_2015}. Downstream processes can be built around any consistent input but not around a variable one.

Reproducibility also underlies heritability.  If trajectory entropy is high, the phenotype varies widely at fixed genotype: the genotype-to-phenotype map becomes unpredictable, phenotypic heritability collapses, and natural selection is ineffective. A mechanism that reduces trajectory entropy restores the mapping from genotype to phenotype, making phenotypic variation selectable.  In our model, the trajectories $x(t)$ realized by a system can be considered the phenotype and the parameters governing the underlying stochastic dynamics --- e.g., molecular interaction parameters that guide self-assembly --- can be considered the genotype.  In this sense, selection for low trajectory entropy is selection for the capacity to undergo Darwinian evolution.

\subsection{Optimal conditioning time}

The optimal reset time $T_r^*$ maximizes $F(T_r)$. Setting $dF/dT_r = 0$:

\begin{equation}
\frac{1}{f_\text{growth}} \frac{df_\text{growth}}{dT_r} + s \frac{d(\Delta S)}{dT_r} = 0.
\label{eq:optimal-Tr}
\end{equation}

The optimum balances two competing effects: loosening the deadline (increasing $T_r$) accepts more trajectories and improves the growth rate, but tightening it (decreasing $T_r$) discards more disordered trajectories and increases entropy reduction.

\subsubsection{Pure growth-rate optimum ($s = 0$)}

At $s = 0$, the optimal $T_r$ maximizes $f_\text{growth}$ alone. Differentiating the Euler-Lotka equation with resets with respect to $T_r$ and setting $df_\text{growth}/dT_r = 0$ gives

\begin{equation}
f_\text{growth}^*(T_r^*) = \frac{P(T_r^*)}{1 - a(T_r^*)},
\label{eq:optimal-hazard}
\end{equation}

which relates the optimal growth rate to the hazard rate of $P(T)$ at $T_r^*$. The optimal reset time is therefore set where the marginal benefit of accepting one more trajectory --- the density $P(T_r^*)$ of trajectories finishing at the deadline --- equals the cost of waiting for it, weighted by the fraction $1 - a(T_r^*)$ of trajectories that would otherwise be lost.

As $s$ increases, $T_r^*$ shifts to smaller values because order has direct fitness value. The critical selection $s_c$ is the smallest $s \geq 0$ at which $\max_{T_r} F(T_r) > f_0$; at $s_c = 0$, conditioning arises spontaneously from growth-rate optimization alone.

\subsection{When does spontaneous conditioning arise?}

Even without selection for entropy reduction ($s = 0$), conditioning can emerge spontaneously if resets increase the growth rate. A well-established result in the stochastic resets literature is that resetting can reduce the mean first-passage time when the completion-time distribution is sufficiently broad \cite{evans_diffusion_2011,evans_stochastic_2020,reuveni_optimal_2016,pal_first_2017}. Our setting is analogous, with the Euler-Lotka growth rate replacing the mean first-passage time. The additional ingredient is that beneficial resets simultaneously reduce trajectory entropy, linking the growth-rate improvement to the emergence of reproducible outcomes.

\subsubsection{Mean-time approximation}

To build intuition for when resets help, consider the simpler question of when they reduce the mean replication time (rather than the full Euler-Lotka growth rate). The mean replication time with resets at $T_r$ satisfies a self-consistency equation: each attempt either finishes before $T_r$ (contributing its completion time) or fails (contributing $T_r$ plus the mean time for all subsequent attempts):

\begin{equation}
\langle T \rangle_{T_r} = \int_0^{T_r} T\, P(T)\, dT + (1 - a(T_r))\left(T_r + \langle T \rangle_{T_r}\right).
\end{equation}

Collecting $\langle T \rangle_{T_r}$ on the left-hand side and dividing by $a(T_r)$ gives

\begin{equation}
\langle T \rangle_{T_r} = \frac{1}{a(T_r)} \left(\int_0^{T_r} T\, P(T)\, dT + (1-a(T_r))\, T_r\right).
\label{eq:mean-time-reset}
\end{equation}

\subsubsection{When do resets speed up replication?}

Resets at time $T_r$ are beneficial only if $\langle T \rangle_{T_r} < E(T)$. Starting from Eq.~\ref{eq:mean-time-reset} and requiring $\langle T \rangle_{T_r} < \int_0^\infty T\, P(T)\, dT$, we multiply both sides by $a(T_r)$:
\begin{eqnarray}
       \int_0^{T_r} T\, P(T)\, dT  &+&  (1-a(T_r))\, T_r   < a(T_r)\int_0^\infty T\, P(T)\, dT\nonumber \\
       &=& \int_0^\infty T\, P(T)\, dT - (1-a(T_r)) \int_0^\infty T\, P(T)\, dT
 \end{eqnarray}

 \begin{eqnarray}
         (1-a(T_r))\, T_r   < \int_{T_r}^\infty T\, P(T)\, dT - (1-a(T_r)) \int_0^\infty T\, P(T)\, dT \nonumber
 \end{eqnarray}

 which simplifies to
\begin{eqnarray}
    \int_{T_r}^{\infty} T\, P(T)\, dT &>& (1-a(T_r))\left( T_r + \int_0^{\infty} T\, P(T)\, dT\right)\\
    E(T | \text{ lost}) &>&  T_r + E(T)\\
    &=&  T_r + \left[ a\, E(T|\text{ not lost})+(1-a)\, E(T|\text{ lost})\right]\nonumber\\
    E(T | \text{ lost}) &>& \frac{T_r}{a(T_r)} + E(T|\text{ not lost})\, . \label{eq:beneficialresets}
\end{eqnarray}
Resets are beneficial when the expected time wasted on a trajectory that would miss the deadline exceeds the cost of the reset ($T_r/a$, accounting for the number of attempts) plus the expected time of a successful attempt. This condition is a mean-time proxy: it gives useful intuition for when discarded trajectories are sufficiently slow.

\subsection{Relation to prior work on stochastic resets}

Stochastic resets have been extensively studied (see \cite{evans_stochastic_2020} for a review), with applications to computer science restart algorithms \cite{luby_optimal_1993,villen-altamirano_restart_nodate}, biological search processes \cite{roldan_stochastic_2016,morris_anillin_2020,rotbart_michaelis-menten_2015}, and animal foraging \cite{viswanathan_optimizing_1999,gordon_local_2011}. Our framework builds on this body of work in several respects. First, we focus on trajectory entropy reduction, beyond growth-rate improvement. Prior work examined the entropy of position distributions or completion-time distributions under resets \cite{fuchs_stochastic_2016,busiello_entropy_2020,eliazar_entropy_2023,louwerse_information_2022}; our interest is in the entropy of paths, which we connect to observable reproducibility in biological products (see also the completion-time entropy proxy defined in Section~\ref{sec:SI-general}). Second, we situate resets in the context of self-replication, where fitness is the Euler-Lotka growth rate rather than a mean first-passage time. This changes the quantitative thresholds for beneficial resets. The renewal approach developed in the resets literature extends naturally to this setting; related extensions include non-Poisson reset-time distributions \cite{eule_non-equilibrium_2016,nagar_diffusion_2016,pal_diffusion_2016}, the inspection paradox \cite{pal_inspection_2022}, search with home returns \cite{pal_search_2020}, and progressive restart strategies \cite{britain_progressive_2022}. Third, we introduce a combined fitness incorporating both growth rate and trajectory entropy reduction, and define a critical selection $s_c$ that quantifies the minimum selective value of entropy reduction at which conditioning becomes favorable. Fourth, we connect temporal conditioning to kinetic proofreading in templated replication (see Section~\ref{sec:SI-physical}), where resetting stalled trajectories is mathematically equivalent to exonuclease-mediated error correction.

\section{The Socks-before-shoes model}
\label{sec:SI-shoessocks}

This section details the socks-before-shoes permutation model introduced in the main text. Instead of continuous trajectories in some high-dimensional space, we define permutations of $n$ discrete actions, where ordering is directly observable and entropy reduction manifests as convergence toward a canonical sequence of those $n$ actions. We can therefore count exactly how much more ordered the conditioned ensemble is compared to the unconditioned one.

\subsection{Model definition}

\subsubsection{The $\epsilon$-model for action selection}

A system must complete $n$ actions, labeled $1, 2, \ldots, n$, in some order. A permutation $\sigma \in S_n$ specifies the sequence: action $\sigma(i)$ is performed at step $i$. The canonical ordering $(1, 2, \ldots, n)$ represents the most efficient sequence.

At each step $i$ (with $n - i + 1$ remaining actions), let $j^*$ denote the smallest-label remaining action. The system selects:

\begin{itemize}
  \item action $j^*$ with probability $(1 - \epsilon) + \epsilon/(n - i + 1)$,
  \item any other remaining action $j \neq j^*$ with probability $\epsilon/(n - i + 1)$.
\end{itemize}

In other words, with probability $1 - \epsilon$ the system follows canonical order, and with probability $\epsilon$ it picks uniformly among all remaining actions. The probability of a full permutation is $P[\sigma] = \prod_{i=1}^{n} p_i(\sigma)$. When $\epsilon = 0$, only the canonical permutation has nonzero probability; when $\epsilon = 1$, all permutations are equally likely.

\subsubsection{Prerequisite-based time costs}

Action $\sigma(i)$ is in order if all actions with smaller canonical labels have already been completed:
\begin{equation}
\{1, 2, \ldots, \sigma(i) - 1\} \subseteq \{\sigma(1), \ldots, \sigma(i-1)\}.
\label{eq:ss-inorder}
\end{equation}
Each in-order step takes time $\tau_\text{fast} = 1$, which sets the unit of time. Each out-of-order step takes a shifted-exponential time
\begin{equation}
\tau_\text{slow} = \tau_\text{fast} + X, \qquad X \sim \text{Exp}\!\left(\text{mean} = (\eta_\text{stall}-1)\,\tau_\text{fast}\right),
\label{eq:ss-shifted-slow-time}
\end{equation}
where the dimensionless \emph{stalling factor} $\eta_\text{stall} = \mathbb{E}[\tau_\text{slow}]/\tau_\text{fast} > 1$ is the mean ratio of slow to fast step times. This convention preserves the intended mean slow-time penalty while ensuring that no out-of-order step can be faster than an in-order step. The total completion time is $T(\sigma) = \sum_{i=1}^{n} \tau_i$. For the canonical permutation, all steps are in order, so $T_\text{canon} = n\,\tau_\text{fast} = n$.

We use prerequisite-based rather than position-match costs because performing an action before its prerequisites (e.g., attaching a component before its neighbors are placed) incurs a real time penalty. A single early error does not cascade: the permutation $(3, 1, 2, 4, 5)$ has only one out-of-order step (the first one, since action 3 was performed before actions 1 and 2).

\subsubsection{Conditioning}

We condition on completion before $T_r$:
\begin{equation}
P^\text{cond}[\sigma] = \frac{P[\sigma] \cdot \Theta(T_r - T(\sigma))}{a(T_r)},
\label{eq:ss-P-cond}
\end{equation}
where $a(T_r) = \sum_\sigma P[\sigma] \cdot \Theta(T_r - T(\sigma))$.

\subsubsection{Computation}
\label{sec:ss-computation}

Two methods are used in this section. (1)~\emph{Analytical method:}  used for all figures except save points. The number of in-order steps in a permutation is a sum of independent Bernoulli indicators (Section~\ref{sec:ss-poisson-binom}), so all observables (acceptance fraction, entropy reduction, growth rate, mean ordering) reduce to closed-form sums over $n+1$ groups, computed in $O(n^2)$ time without sampling (Section~\ref{sec:ss-grouped-EL}). (2)~\emph{Monte Carlo sampling:} used only for the save-point analysis (Section~\ref{sec:ss-savepoints}), where save points couple a trajectory's fate to \emph{which} canonical prefixes were completed and \emph{when}, breaking the group-level structure that the analytical method relies on. Permutations are drawn step by step from the model; with $\tau_\text{fast} = 1$ as the unit, each out-of-order step is assigned an independent shifted-exponential time $1 + \text{Exp}(\text{mean}=\eta_\text{stall}-1)$. Parameters: $N = 200{,}000$ samples, seed 42.

\subsection{Order parameters}

\subsubsection{Ordering}

We quantify temporal order with the ordering score $\kappa(\sigma)$. Let $q(\sigma)$ be the fraction of action pairs that appear in the canonical relative order. Equivalently, let $d_\tau(\sigma, \text{id})$ be the number of pairwise inversions relative to the canonical order. We use the normalized score
\begin{equation}
\kappa(\sigma) = 1 - \frac{2\, d_\tau(\sigma, \text{id})}{\binom{n}{2}},
\label{eq:ss-kendall}
\end{equation}
This is equivalent to $\kappa = 2q - 1$, ranging from $-1$ (fully reversed) to $+1$ (canonical order), with $\kappa = 0$ for a uniformly random permutation on average. We refer to $\kappa$ as the ordering score, defined as the normalized Kendall ordering.

\subsubsection{Pairwise ordering probability}

For each pair $(i, j)$ with $i < j$, the pairwise ordering probability is
\begin{equation}
P_{ij} = \sum_\sigma P[\sigma] \cdot \mathbf{1}[\sigma^{-1}(i) < \sigma^{-1}(j)].
\label{eq:ss-Pij}
\end{equation}

\subsection{Derived thermodynamic quantities}

\subsubsection{Trajectory entropy reduction}

Following the general framework (Eq.~\ref{eq:delta-S}), the trajectory entropy reduction is the Shannon entropy difference
\begin{equation}
\Delta S(T_r) = H[P] - H[P^\text{cond}] = -\sum_\sigma P[\sigma] \ln P[\sigma] + \sum_\sigma P^\text{cond}[\sigma] \ln P^\text{cond}[\sigma],
\label{eq:ss-delta-S}
\end{equation}
computed numerically from the permutation probabilities. Because we work with a discrete set of permutations, the entropy is a finite sum rather than an integral.

\subsubsection{Growth rate with resets}

The growth rate $f_\text{growth}(T_r)$ satisfies
\begin{equation}
2 \sum_\sigma P[\sigma]\, e^{-f_\text{growth}\, T(\sigma)}\, \Theta(T_r - T(\sigma)) = 1 - (1 - a(T_r))\, e^{-f_\text{growth}\, T_r},
\label{eq:ss-euler-lotka}
\end{equation}
solved numerically using Brent's method, a standard root-finding algorithm that reliably converges without requiring derivatives. The unconditioned growth rate $f_0$ satisfies $2 \sum_\sigma P[\sigma]\, e^{-f_0\, T(\sigma)} = 1$.

\subsubsection{Combined fitness and critical selection}

The combined fitness is $F(T_r) = f_\text{growth}(T_r) \cdot e^{s\,\Delta S(T_r)}$ (Eq.~\ref{eq:combined-fitness}). Conditioning is favorable when $F(T_r) > f_0$ for some reset time $T_r$, i.e., when the combined benefit of any growth-rate change and entropy reduction outweighs the unconditioned fitness. Rearranging $F > f_0$ gives the condition on $s$:
\begin{equation}
s > \frac{\ln\bigl(f_0 / f_\text{growth}(T_r)\bigr)}{\Delta S(T_r)}.
\label{eq:ss-sc-condition}
\end{equation}
The right-hand side is the ratio of the growth-rate cost (how much slower the population grows due to resets, expressed as a log-ratio) to the entropy-reduction benefit (how much more reproducible the surviving trajectories are). The critical selection $s_c$ is the smallest value of $s$ for which this inequality can be satisfied at \emph{any} $T_r$. If resets already improve the growth rate at some $T_r$ (so that $f_\text{growth} \geq f_0$), the right-hand side is zero or negative, and $s_c = 0$: conditioning pays for itself through speed alone (synergistic regime). Otherwise, $s_c > 0$: some direct selective value of entropy reduction is needed to offset the growth-rate cost (antagonistic regime).

\subsection{Theory: Poisson Binomial structure and analytical predictions}

\subsubsection{Independence of in-order indicators}

The model is analytically tractable because whether each step is in order is statistically independent of all other steps. At step $i$, the probability that the chosen action is in order is
\begin{equation}
p_i = (1 - \epsilon) + \frac{\epsilon}{n - i}, \qquad i = 0, 1, \ldots, n-1.
\label{eq:ss-pi}
\end{equation}

This probability depends only on the count of remaining actions, not on their identities or history. The reason is simple: the socks-before-shoes model always targets the smallest remaining action with probability $1 - \epsilon$, and picks uniformly among all remaining actions with probability $\epsilon$. Since the remaining set always has the same size $n - i$ at step $i$, the probability of picking the canonical next action is always $p_i$. The indicators $A_i = \mathbf{1}[\text{step } i \text{ is in order}]$ are therefore independent Bernoulli random variables with parameters $p_0, \ldots, p_{n-1}$.

\subsubsection{Poisson Binomial distribution of in-order count}
\label{sec:ss-poisson-binom}

The total number of in-order (io) steps is
\begin{equation}
n_\text{io} = \sum_{i=0}^{n-1} A_i \sim \text{PoissonBinomial}(p_0, \ldots, p_{n-1}).
\label{eq:ss-poisson-binom}
\end{equation}
The Poisson Binomial distribution is the distribution of a sum of independent Bernoulli random variables with \emph{different} success probabilities (unlike the ordinary Binomial, where all probabilities are the same). Its probability mass function $w_k = P(n_\text{io} = k)$ is computed exactly by convolving $n$ Bernoulli PMFs in $O(n^2)$ time, compared to $O(n!)$ for full enumeration of permutations.

\subsubsection{Completion time distribution}

Since every step contributes one unit of base time and each of the $n-k$ out-of-order steps contributes an independent exponential extra delay with mean $\eta_\text{stall}-1$ (with $\tau_\text{fast}=1$), the completion time for a permutation with $n_\text{io}=k$ in-order steps is
\begin{equation}
T \mid n_\text{io} = k \;\sim\; n + \text{Gamma}(n-k,\; \text{scale} = \eta_\text{stall}-1).
\label{eq:ss-T-gamma}
\end{equation}
The mean completion time for group $k$ is $n + (n-k)(\eta_\text{stall}-1) = k + (n-k)\eta_\text{stall}$, which is strictly decreasing in $k$ since $\eta_\text{stall} > 1$: more in-order steps means faster completion. Thus the shifted-exponential convention leaves the mean-time parametrization unchanged, but enforces the hard lower bound $T \ge n$. The probability that such a permutation completes before a deadline $T_r$ is therefore
\begin{equation}
z_k(T_r) = F_\text{Gamma}(T_r - n;\; n-k,\; \eta_\text{stall}-1),
\label{eq:ss-zk}
\end{equation}
where $F_\text{Gamma}$ is the Gamma cumulative distribution function, with the convention $z_k(T_r)=0$ for $T_r<n$ and $z_n(T_r)=\mathbf{1}[T_r\ge n]$. The total acceptance fraction is then $a(T_r) = \sum_{k=0}^{n} w_k\, z_k(T_r)$.

\subsubsection{Grouped Euler-Lotka equation}
\label{sec:ss-grouped-EL}

Because all permutations with the same number of in-order steps $k$ have the same time statistics, we can replace the sum over all $n!$ permutations with a sum over $n+1$ groups. For each group $k$, we define the Laplace-weighted completion probability in terms of the extra delay $u = T-n$:
\begin{equation}
L_k(T_r, f_\text{growth}) = e^{-f_\text{growth} n}\int_0^{T_r - n} e^{-f_\text{growth}\,u}\, g(u;\, n-k,\, \eta_\text{stall}-1)\, du,
\label{eq:ss-laplace-integral}
\end{equation}
where $g(\cdot;\, \alpha, \eta_\text{stall}-1)$ is the Gamma density with shape $\alpha = n - k$ and scale $\eta_\text{stall}-1$. For the all-in-order group $k=n$, the extra delay is identically zero, so $L_n(T_r,f_\text{growth})=e^{-f_\text{growth}n}\mathbf{1}[T_r\ge n]$. Using the known Laplace transform of the Gamma distribution for $k<n$, this evaluates to
\begin{equation}
L_k(T_r, f_\text{growth}) = e^{-f_\text{growth} n}\,\frac{1}{\left[1 + f_\text{growth}\,(\eta_\text{stall}-1)\right]^{n-k}}\; F_\text{Gamma}\!\left(T_r - n;\; n-k,\; \frac{\eta_\text{stall}-1}{1+f_\text{growth}\,(\eta_\text{stall}-1)}\right).
\label{eq:ss-laplace-gamma}
\end{equation}
The Euler-Lotka equation with resets at $T_r$ then becomes a finite sum over groups:
\begin{equation}
2 \sum_{k=0}^{n} w_k\, L_k(T_r, f_\text{growth}) = 1 - (1 - a(T_r))\, e^{-f_\text{growth}\, T_r},
\label{eq:ss-euler-lotka-continuous}
\end{equation}
where $a(T_r) = \sum_k w_k\, z_k(T_r)$. Without resets ($T_r \to \infty$), $L_k \to e^{-f_\text{growth}n}\left[1+f_\text{growth}(\eta_\text{stall}-1)\right]^{-(n-k)}$ and the equation simplifies to $2 e^{-f_\text{growth}n}\sum_k w_k\, \left[1+f_\text{growth}(\eta_\text{stall}-1)\right]^{-(n-k)} = 1$.

\label{sec:ss-eps-c}
The same grouped computation determines the synergistic boundary $\epsilon_c(\eta_\text{stall})$, the largest $\epsilon$ at which $s_c = 0$. For each candidate $\epsilon$, we compute the Poisson Binomial weights $w_k$, solve for the baseline growth rate $f_0$ (without resets), and scan over $T_r$ to check whether any reset time gives $f_\text{growth}(T_r) > f_0$. A binary search over $\epsilon$ locates the boundary $\epsilon_c$.

\subsection{Simulation parameters}

\begin{figure*}
  \includegraphics[width=\textwidth]{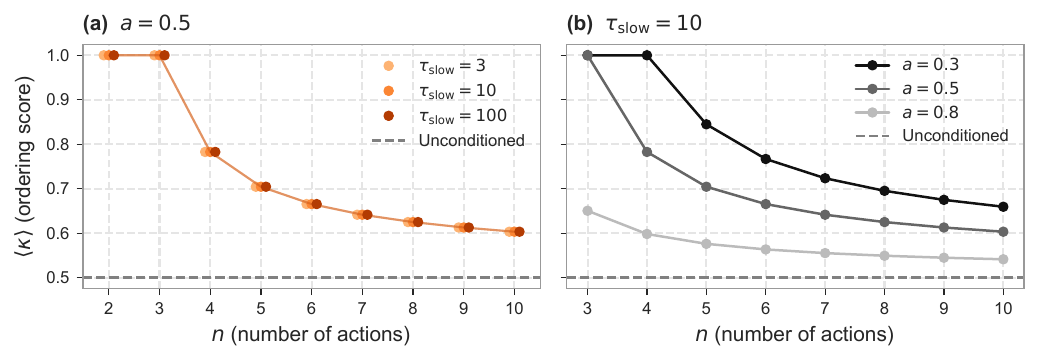}
  \caption{Conditioned mean ordering score $\langle\kappa\rangle$ versus number of actions $n$ ($\epsilon = 0.5$). All computations use an analytical method (shifted Gamma-CDF summation over $n+1$ groups, no Monte Carlo). The dashed gray line marks the unconditioned baseline $\langle\kappa\rangle = 1 - \epsilon = 0.5$. \textbf{(a)}~ Fixed acceptance $a = 0.5$, varying stalling factor $\eta_\text{stall} \in \{3, 10, 100\}$. The three curves with different $\tau_\text{slow}$ are identical because they are being compared at a fixed acceptance $a$. \textbf{(b)}~ Fixed $\eta_\text{stall} = 10$, varying acceptance $a \in \{0.3, 0.5, 0.8\}$ ($n \geq 3$). Stricter conditioning (smaller $a$) consistently produces more ordering across all $n$, with the gap between curves remaining roughly constant as $n$ grows.}
  \label{fig:SI_scaling}
\end{figure*}

All panels of the main-text figures, except for the savepoint panels, use the analytical method (Poisson Binomial weights $w_k$ combined with Gamma-CDF completion probabilities, Eqs.~\ref{eq:ss-zk}--\ref{eq:ss-euler-lotka-continuous}). The save-point figure (main text) uses Monte Carlo sampling as described in Section~\ref{sec:ss-computation}. Parameters for each panel are listed in Table~\ref{tab:ss-params}. Fig.~\ref{fig:SI_scaling} shows how the conditioned mean ordering scales with $n$ for different values of the stalling factor $\eta_\text{stall}$ and acceptance fraction $a$, extending the main-text results to a wider range of system sizes.

\begin{table}[h]
\centering
\begin{tabular}{l l l l l l}
\hline
Figure & Panel & $n$ & $\eta_\text{stall}$ & $\epsilon$ & Method / Notes \\
\hline
Main Fig.~\ref{fig:shoessocks} & (a) & 8 & 100 & 0.01, 0.1, 0.5 & Analytical, 400 log-spaced $T_r/T_\text{min}$ \\
Main Fig.~\ref{fig:shoessocks} & (b) & 8 & 3 & 0.5 & Analytical, $a \approx 0.95, 0.50, 0.20$ \\
Main Fig.~\ref{fig:shoessocks} & (c) & 8 & 100 & 0.1, 0.5,0.6 & Analytical, 400 log-spaced $T_r$ \\
Main Fig.~\ref{fig:shoessocks} & (d) & 8 & 3, 10, 30, 100 & 0.01--1.0 (50 pts) & Analytical \\
Main Fig.~\ref{fig:savepoints} & -- & 4, 6, 8, 10 & 50 & 0.5 & Monte Carlo, $N = 200{,}000$, seed 42 \\
Fig.~\ref{fig:SI_scaling} & -- & 2--10 & 3, 10, 100 & 0.5 & Analytical \\
Fig.~\ref{fig:SI_synergistic} & -- & 8 & 3--300 (7 pts) & 0.01--0.5 (15 pts) & Analytical \\
\hline
\end{tabular}
\caption{Parameters for all socks-before-shoes model figures.}
\label{tab:ss-params}
\end{table}

\subsection{The antagonistic-to-synergistic transition}

The transition is controlled by the stalling factor $\eta_\text{stall} = \tau_\text{slow}/\tau_\text{fast}$. At small $\eta_\text{stall}$ (e.g., $\eta_\text{stall} = 3$), out-of-order steps are only moderately slower than in-order ones, and the time spent on a failed attempt before resetting is largely wasted (antagonistic). At large $\eta_\text{stall}$ (e.g., $\eta_\text{stall} = 100$), even a single out-of-order step makes a trajectory dramatically slower than the canonical one, so resetting early and trying again is faster than finishing a disordered trajectory (synergistic). This parallels the role of tail weight in the distribution family survey: small $\eta_\text{stall}$ produces a completion-time distribution with a modest right tail, while large $\eta_\text{stall}$ produces a heavy tail where the slowest trajectories are orders of magnitude slower than the fastest.

\begin{figure}
  \includegraphics[width=0.4\columnwidth]{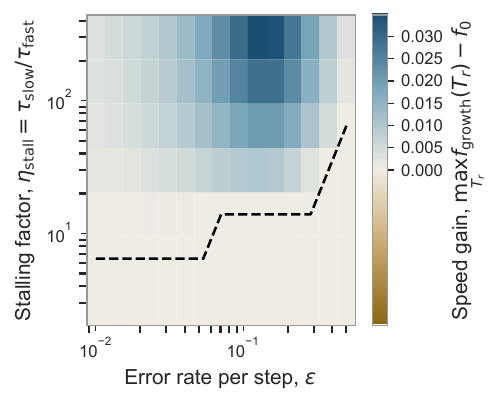}
  \caption{Synergistic regime in the socks-before-shoes model. Speed gain from resets, $\max_{T_r} f_\text{growth}(T_r) - f_0$, across the $(\epsilon, \eta_\text{stall})$ parameter space for $n = 8$ actions. Blue cells indicate the synergistic regime where resets improve the growth rate ($f_\text{growth} > f_0$); gold cells indicate the antagonistic regime where resets always reduce speed. The dashed contour marks the boundary between regimes, $\max_{T_r} f_\text{growth}(T_r) = f_0$, and is numerically determined from Eq.~\ref{eq:ss-euler-lotka-continuous}. At small $\eta_\text{stall}$ (e.g., $\eta_\text{stall} = 3$), the model is purely antagonistic for all $\epsilon$, consistent with Fig.~\ref{fig:shoessocks}(d). At large $\eta_\text{stall}$ (roughly $\eta_\text{stall} \gtrsim 15$ for small $\epsilon$), a synergistic regime emerges: resetting before accumulating too many slow out-of-order steps is faster than completing them. The synergistic region expands toward larger $\epsilon$ as $\eta_\text{stall}$ increases. $\epsilon$ ranges from 0.01 to 0.5 in 15 log-spaced steps; $\eta_\text{stall}$ ranges from 3 to 300 in 7 log-spaced steps. Computed using the analytical method over $n+1$ groups for $n = 8$.}
  \label{fig:SI_synergistic}
\end{figure}

Fig.~\ref{fig:SI_synergistic} maps the growth-rate gain $\max_{T_r} f_\text{growth}(T_r) - f_0$ across the $(\epsilon, \eta_\text{stall})$ parameter space for $n = 8$. The synergistic region enters at roughly $\eta_\text{stall} \approx 15$ for small $\epsilon$ and expands toward larger $\epsilon$ as $\eta_\text{stall}$ increases.

\subsection{Save points: partial resets from intermediate states}
\label{sec:ss-savepoints}

\subsubsection{Motivation and model}

For large $n$, conditioning becomes antagonistic because the combinatorial explosion of the permutation space makes it too costly to discard all slow trajectories. Save points address this problem by decomposing the large problem into smaller sub-problems: positions $k$ in the canonical sequence at which verified progress is preserved across resets. If a trajectory has completed actions $1, \ldots, k$ in order, a reset returns to position $k$ rather than to the start.

Let $\mathcal{K} = \{k_1, \ldots, k_m\}$ be the save-point positions and let $S_j$ denote the state in which the canonical prefix $1, 2, \ldots, k_j$ has been completed in order. Starting from $S_j$, the system attempts the residual problem on actions $k_j+1, \ldots, n$. Within one reset period $T_r$, three outcomes are possible: (i) complete the entire task before $T_r$ (probability $A_j$); (ii) reach some higher save point $S_\ell$ ($\ell > j$) but fail to complete before $T_r$ (probability $B_{j \to \ell}$, so that the next attempt starts from $S_\ell$); or (iii) fail to reach any higher save point and reset back to $S_j$ (probability $C_j = 1 - A_j - \sum_{\ell > j} B_{j\to\ell}$). After such a reset, the next attempt starts from the new save point with a fresh reset window of duration $T_r$ applied to the residual problem.

\subsubsection{Euler-Lotka equation with save points}

Each state $S_j$ contributes to the population growth through a generating function $G_j(f_\text{growth})$ that accounts for all possible futures starting from that state:
\begin{equation}
G_j(f_\text{growth}) = \frac{A_j(f_\text{growth}) + \sum_{\ell > j} B_{j \to \ell}\, e^{-f_\text{growth}\,T_r}\, G_\ell(f_\text{growth})}{1 - C_j\, e^{-f_\text{growth}\,T_r}},
\label{eq:ss-savepoint-G}
\end{equation}
where $A_j(f_\text{growth}) = \sum_{\sigma \in \text{complete}} P[\sigma]\, e^{-f_\text{growth}\, T(\sigma)}$ is the Laplace-weighted acceptance from $S_j$. The numerator has two terms: either the system completes from this state (weighted by $A_j$), or it advances to a higher save point and then completes from there. The denominator accounts for repeated failures at the same save point, each costing time $T_r$. The growth rate satisfies $2\, G_0(f_\text{growth}) = 1$, solved by backward substitution starting from the highest save point and Brent root-finding.

\subsubsection{Ordering}

The steady-state $\langle\kappa\rangle$ is computed by weighting each state's contribution by the probability flow $\phi_j$ through that state ($\phi_0 = 1$, $\phi_\ell = \sum_{j < \ell} \phi_j\, B_{j \to \ell}/(1 - C_j)$). Conditional on being in $S_j$, the $\binom{k_j}{2}$ pairs entirely within the prefix are correctly ordered (every smaller-label action precedes every larger one), and the $k_j(n - k_j)$ pairs that link a prefix action to a residual action $j' > k_j$ are also correctly ordered (the prefix action was performed first); together these contribute the $\binom{k_j}{2} + k_j(n - k_j)$ deterministic correctly ordered pairs. The remaining $\binom{n - k_j}{2}$ pairs lie entirely within the residual block and their ordering is averaged over the conditioned residual ensemble (the $\epsilon$-model restricted to actions $k_j+1, \ldots, n$, conditioned on completion before $T_r$).

\subsubsection{Computation}
\label{sec:ss-savepoint-computation}

The Poisson Binomial group structure that drives the analytical method elsewhere in this section breaks down here, because save-point transitions depend on \emph{which} canonical prefix has been completed and on the time at which it was completed, not just on the number of in-order steps. We use a two-layer scheme in which Monte Carlo handles only the parts that do not allow a closed-form treatment, while the Euler-Lotka equation and the ordering average are still computed analytically:

\begin{enumerate}
\item \emph{Sampling layer.} For each starting state $S_j$, we draw $N = 200{,}000$ samples (seed 42) from the $\epsilon$-model on the residual actions, with shifted-exponential slow-step times $1+\text{Exp}(\text{mean}=\eta_\text{stall}-1)$ in units where $\tau_\text{fast}=1$. From these samples we estimate, at the chosen reset time $T_r$, the transition probabilities $A_j$, $B_{j\to\ell}$, $C_j$, the Laplace-weighted acceptance $A_j(f_\text{growth}) = \sum_{\sigma \in \text{complete}} P[\sigma]\, e^{-f_\text{growth}\,T(\sigma)}$, and the residual-block conditional ordering score $\kappa$ at each state.
\item \emph{Analytical layer.} Given those coefficients, Eq.~\ref{eq:ss-savepoint-G} is solved by backward substitution from the highest save point, $f_\text{growth}$ is extracted from $2\,G_0(f_\text{growth}) = 1$ via Brent's method, and $\langle\kappa\rangle$ is assembled by weighting each state's prefix/residual decomposition by the probability flow $\phi_j$. The randomness in the final answer therefore comes only from the coefficient estimates, not from any explicit sampling of the growth rate.
\end{enumerate}

A deterministic-time mode of the same code in which slow-step times are pinned to their mean $\eta_\text{stall}\,\tau_\text{fast}$ and all $n!$ permutations are enumerated, computes the same coefficients exactly and is used as a cross-check at small $n$.

\section{Distribution family survey}
\label{sec:SI-distributions}

\subsection{Overview}

In the socks-before-shoes model (Section~\ref{sec:SI-shoessocks}), the completion-time distribution emerges from microscopic parameters ($\epsilon$, $\eta_\text{stall}$) that may be difficult to measure directly. In many experimental settings, however, the completion-time distribution itself is the primary observable: cell-division times, folding times, or replication times can be measured without detailed knowledge of the underlying kinetics. This motivates a complementary, phenomenological approach in which we take the completion-time distribution $P(T)$ as given and ask how its shape determines whether conditioning is favorable.

The main text shows that the shape of $P(T)$ determines whether conditioning through resets is synergistic (improving both growth rate and order) or antagonistic (trading off one for the other). To understand how this depends on the distribution, we evaluate the conditioning framework across five standard families: Normal, Weibull, Log-normal, Fr\'{e}chet, and Gumbel. These families span the full spectrum of tail behavior, from distributions where extreme values are rare and close to the mean (Normal), to distributions with power-law tails where occasional events can be vastly slower than typical (Fr\'{e}chet). This range lets us identify what feature of the distribution controls the transition.

For the strictly positive families (Weibull, Log-normal, and Fr\'{e}chet), we fix the mean to unity so that differences reflect shape rather than overall timescale. For Normal and Gumbel, which have support on the full real line, we set the raw full-line distribution to have mean one and then condition on positive completion times ($T>0$); after this truncation and renormalization, the effective positive-time distribution generally has mean greater than one. This convention preserves the raw mean-one parameterization while enforcing physical positive completion times. We then vary a single parameter that controls the width or tail weight. Throughout this section, $\Delta S$ denotes the completion-time Shannon entropy difference $H_T[P]-H_T[P^\text{cond}]$, used as the distribution-level proxy for ordering; it is non-negative over the plotted parameter ranges.

\begin{figure*}
  \includegraphics[width=0.78\textwidth]{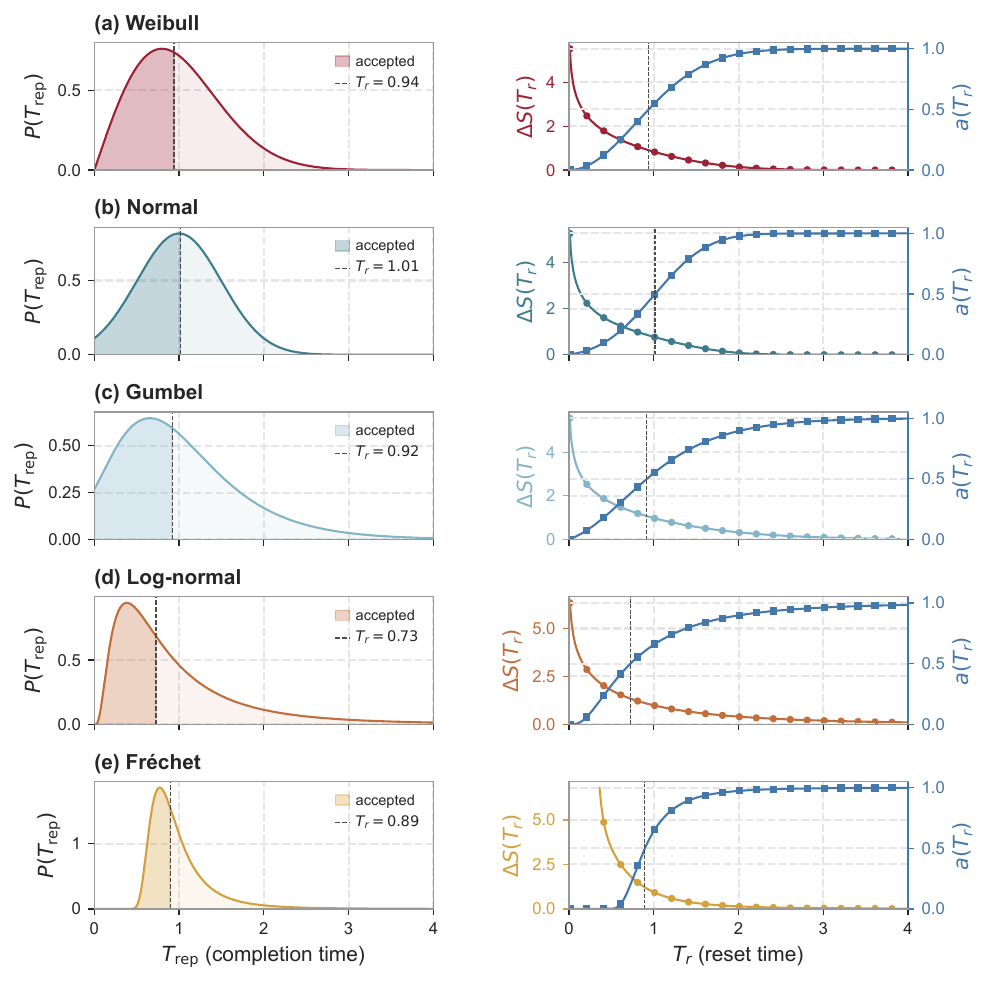}
  \caption{\textbf{Conditioning across the five distribution families.} Each row shows one family at a representative shape parameter, with rows ordered by tail heaviness from narrow (top) to heavy-tailed (bottom): (a) Weibull ($k = 2$, super-exponential decay), (b) Normal ($\sigma = 0.5$), (c) Gumbel ($\beta = 0.6$), (d) Log-normal ($\sigma = 0.8$), (e) Fr\'{e}chet ($\alpha = 4$, power-law tail). Weibull, Log-normal, and Fr\'{e}chet are scaled to unit mean; Normal and Gumbel are raw mean-one full-line distributions conditioned on $T>0$ and renormalized. \emph{Left column.} PDF $P(T_\text{rep})$ with the reset cutoff $T_r$ drawn as a vertical dashed line. The cutoff in each row is set so that the acceptance fraction $a(T_r) = 0.5$ (the median); the resulting $T_r$ values vary across families because the positive-time distribution depends on its shape and, for Normal and Gumbel, on the effect of truncation. The shaded region under the curve marks the accepted trajectories. \emph{Right column.} Shannon entropy reduction $\Delta S(T_r) = H[P] - H[P^\text{cond}]$ (left axis, family color) and acceptance fraction $a(T_r)$ (right axis, blue) vs the reset time. The dashed vertical line marks the cutoff used in the left panel and crosses the acceptance curve at $a = 0.5$ by construction. In all five panels, $\Delta S$ decreases monotonically with $T_r$ and $a$ increases monotonically; only the magnitudes differ. Heavier-tailed families achieve larger $\Delta S$ than narrower families, reflecting the larger entropy reservoir of their tails.}
  \label{fig:SI_concept_families}
\end{figure*}

Before the family-by-family analysis, Fig.~\ref{fig:SI_concept_families} illustrates how the conditioning framework introduced in main-text Fig.~1 generalizes across all five families. For each family at a representative shape parameter, we set the truncation cutoff at the median of the implemented positive-time distribution $P(T)$ (acceptance fraction $a(T_r) = 0.5$) and plot the conditioned ensemble alongside $\Delta S(T_r)$ and $a(T_r)$ as functions of the reset time. The same qualitative pattern appears in these representative examples: earlier reset times narrow the accepted completion-time ensemble and increase the completion-time entropy reduction, with only the magnitude varying across families. In the plotted parameter choices, the maximum completion-time entropy reductions are of order $5$--$7$ nats, with the heavier-tailed Log-normal and Fr\'{e}chet rows larger than the narrower rows.

The numerical procedure for each parameter value is:

\begin{enumerate}
\item Solve the Euler-Lotka equation with resets (Eq.~\ref{eq:euler-lotka-reset}) for $f_\text{growth}(T_r)$ on a grid of $T_r$ values. This is done by root-finding using Brent's method.
\item Compute the entropy reduction $\Delta S(T_r)$ and the combined fitness $F(T_r) = f_\text{growth}(T_r) \cdot e^{s\,\Delta S}$ for a range of $s$. Optimize over $T_r$ to obtain $T_r^*(s)$, $\Delta S^*(s)$, and $f_\text{growth}^*(s)$.
\item Determine $s_c$ by binary search over $s$: the smallest $s \in [0,1)$ at which $\max_{T_r} F(T_r) > f_0$.
\end{enumerate}

In the numerical implementation, $f_0$, $f_\text{growth}(T_r)$, and the completion-time entropy reduction are evaluated on deterministic quadrature grids and solved by Brent root-finding. The same grid-based method is used across families in the distribution survey so that relative comparisons are computed consistently. The baseline fitness $f_0$ is obtained from $2\tilde{P}(-f_0) = 1$ without resets.

\subsection{Distribution definitions and parameterizations}

\subsubsection{Normal}

\begin{equation}
P(\tau) = \frac{1}{Z_N\,\sigma\sqrt{2\pi}} \exp\!\left(-\frac{(\tau - 1)^2}{2\sigma^2}\right), \qquad \tau>0,
\label{eq:normal-pdf}
\end{equation}
where $Z_N = 1 - \Phi(-1/\sigma)$ is the positive-time normalization. The raw Gaussian has mean one before truncation; under the positive-truncated convention, the effective mean is $E[T\mid T>0] > 1$ when the negative tail is appreciable.

Width parameter: $\sigma \in [0.05, 1.80]$ (truncated at $\tau = 0$ and renormalized). The variance reported in the survey is the variance of the truncated positive-time distribution, not the raw full-line value $\sigma^2$. Extreme completion times are exponentially suppressed.

The acceptance fraction is evaluated in closed form:

\begin{equation}
a(T_r) = \frac{\Phi\!\left((T_r-1)/\sigma\right)-\Phi\!\left(-1/\sigma\right)}{1-\Phi\!\left(-1/\sigma\right)}.
\label{eq:normal-acceptance}
\end{equation}

The corresponding Laplace-weighted integral is evaluated numerically on the shared quadrature grid used for all families.

Resets never increase growth rate for any $\sigma$ in the tested range: the thin tails ensure that lost trajectories are only marginally slower than accepted ones, so resetting them does not save enough time to compensate for the cost of the reset. Consequently, $s_c > 0$ throughout, decreasing slowly with $\sigma$.

\subsubsection{Weibull}

\begin{equation}
P(\tau) = k\,\Gamma(1+1/k)\,\bigl[\tau\,\Gamma(1+1/k)\bigr]^{k-1}\,\exp\!\bigl[-\bigl(\tau\,\Gamma(1+1/k)\bigr)^k\bigr].
\label{eq:weibull-pdf}
\end{equation}

Mean-fixing: $\lambda_W = 1/\Gamma(1+1/k)$. Shape parameter: $k \in [0.50, 3.00]$. The Weibull family interpolates between tail behaviors: stretched-exponential tails for $k > 1$, exponential at $k = 1$, and heavy-tailed for $k < 1$. Stretched-exponential ($k < 1$) tails arise generically when relaxation is governed by a broad spectrum of timescales, as in glassy and rugged-energy-landscape dynamics. The transition from $s_c > 0$ to $s_c = 0$ occurs as the Weibull crosses from light-tailed to heavy-tailed behavior.

\subsubsection{Log-normal}

\begin{equation}
P(\tau) = \frac{1}{\tau \sigma \sqrt{2\pi}} \exp\!\left(-\frac{(\ln \tau + \sigma^2/2)^2}{2\sigma^2}\right).
\label{eq:lognormal-pdf}
\end{equation}

Mean-fixing: $\mu_\text{LN} = -\sigma^2/2$. Width parameter: $\sigma \in [0.30, 1.40]$. Variance: $e^{\sigma^2} - 1$.

\subsubsection{Fr\'{e}chet}

\begin{equation}
P(\tau) = \frac{\alpha}{\beta}\left(\frac{\tau}{\beta}\right)^{-(\alpha+1)} \exp\!\left[-\left(\frac{\tau}{\beta}\right)^{-\alpha}\right], \qquad \tau > 0.
\label{eq:frechet-pdf}
\end{equation}

\begin{figure*}
  \includegraphics[width=\textwidth]{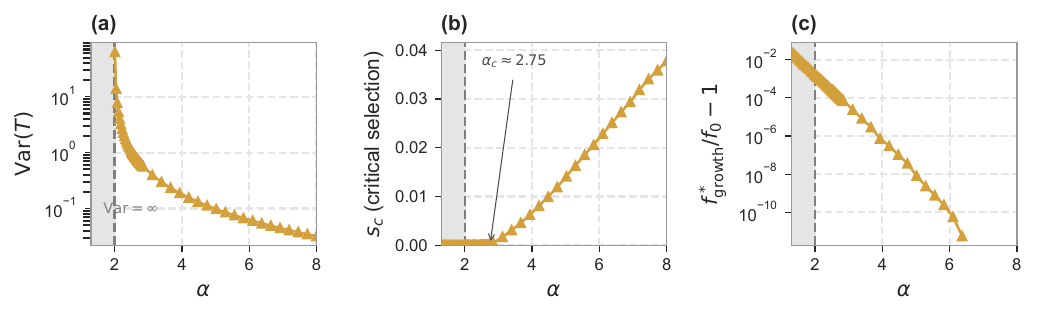}
  \caption{Fr\'{e}chet distribution: full-range analysis across tail exponent $\alpha$. (a) Completion-time variance $\mathrm{Var}(T)$ vs $\alpha$. The variance diverges at $\alpha = 2$ (dashed line; shaded region indicates $\mathrm{Var} = \infty$) and decreases monotonically for $\alpha > 2$. (b) Critical selection $s_c$ vs $\alpha$. For $\alpha < \alpha_c \approx 2.75$, $s_c = 0$: conditioning is spontaneously favorable as a byproduct of growth-rate optimization alone. Above $\alpha_c$, $s_c$ increases steadily as the tail lightens. The transition occurs at finite variance ($\mathrm{Var} \approx 0.68$), well above the $\alpha = 2$ divergence boundary. (c) Optimal growth-rate fold-change $f_\text{growth}^*/f_0 - 1$ vs $\alpha$ (log scale). The fold-change exceeds zero for $\alpha < \alpha_c$ (resets increase growth rate) and approaches zero from above for larger $\alpha$. In the infinite-variance regime ($\alpha \leq 2$), the fold-change grows further, reaching $\approx 0.03$ at $\alpha = 1.3$. All distributions are normalized to unit mean with $\beta = 1/\Gamma(1 - 1/\alpha)$.}
  \label{fig:SI_frechet}
\end{figure*}

\begin{figure}
  \includegraphics[width=0.4\columnwidth]{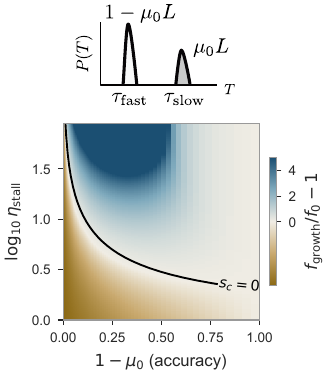}
  \caption{Two-delta model of templated replication with stalling. Fold-change in replication speed ($f_\text{growth}/f_0 - 1$) with optimal resets as a function of accuracy ($1-\mu_0$, x-axis) and stalling ratio ($\log_{10}\eta_\text{stall}$, y-axis), where $\eta_\text{stall} = \tau_s/\tau_f$. A fraction $\mu$ of replication events stall (completion time $\tau_s$), while fraction $1-\mu$ proceed without errors (time $\tau_f$). The black contour marks the mean-time proxy boundary for speed-favourable proofreading: above this guide line, resets are expected to increase speed by the mean-time criterion; below it, resets require direct fitness value of entropy reduction. The boundary diverges as $\mu \to 1$, reflecting that aggressive resets require proportionally larger stalling.}
  \label{fig:SI_twodelta}
\end{figure}

Mean-fixing: $\beta = 1/\Gamma(1 - 1/\alpha)$. Shape parameter: $\alpha \in [2.50, 8.00]$ for the main survey; Fig.~\ref{fig:SI_frechet} extends to $\alpha = 1.3$ to include the infinite-variance regime. The mean is finite only for $\alpha > 1$, so the unit-mean rescaling itself requires $\alpha > 1$; the framework is well-defined throughout the range plotted in Fig.~\ref{fig:SI_frechet} ($\alpha \geq 1.3$). The variance is finite for $\alpha > 2$ and diverges for $\alpha \leq 2$, reflecting the power-law tail $P(\tau) \sim \tau^{-(\alpha+1)}$. The Fr\'{e}chet distribution is the extreme-value distribution for maxima of heavy-tailed data.

The $s_c = 0$ transition at $\alpha_c \approx 2.74$ occurs at finite variance ($\mathrm{Var}(T) \approx 0.68$), so infinite variance is not required for spontaneous conditioning; what matters is the tail exponent. Main-text Fig.~\ref{fig:survey} plots $s_c$ vs.\ variance and therefore shows only the finite-variance portion ($\alpha > 2$); Fig.~\ref{fig:SI_frechet} shows $s_c$, $\mathrm{Var}(T)$, and growth-rate fold-change as functions of $\alpha$ directly, including the infinite-variance regime. For $\alpha \leq 2$, $s_c$ remains zero and the growth-rate advantage persists; the conditioning framework is well-defined because $T_r$ truncates the tail, keeping $a(T_r)$ and $\Delta S$ finite.

\subsubsection{Gumbel}

\begin{equation}
P(\tau) = \frac{1}{Z_G\,\beta} \exp\!\left[-\frac{\tau - (1 - \beta\gamma)}{\beta} - \exp\!\left(-\frac{\tau - (1 - \beta\gamma)}{\beta}\right)\right], \qquad \tau>0.
\label{eq:gumbel-pdf}
\end{equation}

Raw-location convention: $\mu_G = 1 - \beta\gamma$, where $\gamma \approx 0.5772$ is the Euler-Mascheroni constant, so the untruncated full-line Gumbel has mean one. Under the positive-truncated convention, the implemented distribution is truncated at $\tau=0$ and renormalized by $Z_G = 1 - F_G(0)$; its effective mean is generally greater than one. Scale parameter: $\beta \in [0.15, 2.50]$. The variance reported in the survey is the variance of this truncated positive-time distribution. The Gumbel is the extreme-value distribution for maxima of thin-tailed data, with an exponential right tail. Resets never improve the growth rate in the tested range ($\beta \leq 2.5$): $s_c > 0$ throughout, decreasing steadily with $\beta$ but remaining finite.

\section{Physical conditioning}
\label{sec:SI-physical}

The main text treats conditioning as a mathematical operation: restricting the probability distribution over trajectories to those that complete by time $T_r$. Here we emphasize that in physical and biological systems, this mathematical conditioning must be realized by a physical mechanism, and that this requirement has concrete implications.

\subsection{Products encode trajectories}

Stochastic processes in biology produce physical objects: a partially assembled microtubule, a folded or misfolded protein, a newly synthesized strand of DNA. These objects persist and carry information about the trajectories that created them. The length of a microtubule records how many tubulin subunits were added and in what structural register, the conformation of a misfolded protein reflects the kinetic trap its folding pathway encountered, and a mismatch in a DNA strand marks where a polymerase erred. In this sense, the products of stochastic processes are fossils of their trajectories: physical records that persist after the process is complete.

For the mathematical conditioning described in the main text to produce physical order, it is not enough to restrict attention to successful trajectories in a theoretical ensemble. The products of failed trajectories must be physically destroyed, so that only products of the conditioned ensemble exist in the cell. Without this destruction, the physical population of products would reflect the unconditioned distribution, regardless of what the mathematical conditioning predicts. Physical conditioning therefore requires active degradation, disassembly, or displacement of the products of trajectories that fail to meet the criterion.

\subsection{Biological mechanisms of physical conditioning}

Biology implements this physical destruction in many ways. We sketch a few examples in which the mapping to conditioning through resets is particularly direct.

A clear case is the dynamic instability of microtubules \cite{mitchison_dynamic_1984}. Microtubules alternate between growth and rapid depolymerization (catastrophe), and a filament that has not been stabilized by binding its target (e.g., a kinetochore) within a characteristic time depolymerizes back to the centrosome. The incomplete trajectory is erased and its tubulin subunits returned to the pool for a new attempt. No molecular recognition of a specific error is needed: the clock is essentially set by the GTP hydrolysis rate in the microtubule lattice \cite{holy_dynamic_1994}.

Proteostasis offers a similar story at the molecular scale. Proteins that fail to fold within the time window provided by the endoplasmic reticulum's chaperone machinery are retrotranslocated to the cytoplasm and degraded by the proteasome \cite{vembar_one_2008}, breaking the misfolded protein into peptides that retain no memory of the failed folding pathway. Messenger RNAs tell a parallel story: transcripts carrying premature stop codons or other defects are recognized and destroyed by the nonsense-mediated decay pathway \cite{kervestin_nmd_2012} before the error can propagate to the protein level, while the released ribosomes and nucleotides become available for new rounds of translation.

Mismatch repair is a useful intermediate case. When replication introduces a mismatched base pair, the repair system excises a stretch of the newly synthesized strand and resynthesizes it \cite{modrich_mechanisms_2006}. Repair is directional --- it targets the new strand, not the template --- so the record of the correct trajectory is preserved while the record of the failed one is erased.

The most dramatic example is developmental apoptosis, where the entire cell is the physical record of its trajectory. Cells that fail to receive survival signals within a time window undergo programmed cell death \cite{fuchs_programmed_2011}, and the released amino acids, nucleotides, and lipids are recycled by neighbors. An entire cellular trajectory is eliminated on the basis of a temporal signal.

\subsection{Connection to the resets framework}

In each of these examples, the physical mechanism maps directly onto the reset operation in our formalism. The destruction of the product corresponds to the restart of the process at $t = 0$. The time scale of the destruction mechanism (GTP hydrolysis rate, ER retention time, mRNA half-life, cell-cycle checkpoint timing) sets the reset time $T_r$. The cost of resetting, which appears in the compound fitness $\ln F = \ln f_\text{growth} + s\,\Delta S$, corresponds to the physical cost of degradation and disassembly --- ATP spent on proteolysis, tubulin and nucleotides that must be resupplied for fresh attempts, and so on.

The clock-based nature of conditioning (completion time as the sole criterion, with no recognition of the specific error) maps onto the indiscriminate character of many of these mechanisms. Dynamic instability does not identify which tubulin subunit is misplaced; it simply depolymerizes the entire structure if stabilization has not occurred. ER-associated degradation does not diagnose the specific misfolded domain; it retrotranslocates the entire protein. This indiscriminate destruction is what makes conditioning mechanistically simpler than specific error correction, requiring only a clock rather than a high-dimensional molecular recognition apparatus.

\section{Application to templated replication}
\label{sec:SI-applications}

\subsection{Biological context: stalling in templated replication}

Templated replication, in which a polymer is synthesized by reading an existing template strand, is a universal feature of living systems. DNA polymerases copy chromosomes with remarkable fidelity \cite{bustamante_revisiting_2011}, with per-nucleotide error rates $\mu \sim 10^{-8}$--$10^{-9}$. RNA polymerases have higher error rates, $\mu \sim 10^{-4}$--$10^{-5}$. At the other extreme, prebiotic replication systems such as ribozymes and non-enzymatic template-directed synthesis \cite{rajamani_effect_2010} operate near the physicochemical limit, with $\mu \sim 10^{-1}$--$10^{-2}$. Across all of these systems, a common kinetic feature accompanies misincorporation: stalling. When a polymerase incorporates the wrong nucleotide, the resulting mispair distorts the active site geometry, slowing or halting further extension. The severity of this effect varies enormously. In high-fidelity DNA polymerases, a single mismatch can reduce the extension rate by factors of $10^2$--$10^4$, effectively arresting the polymerase until an exonuclease removes the error. In non-enzymatic replication, stalling is more modest but still measurable: mismatches slow primer extension by factors of $\sim$2--10 in non-enzymatic RNA copying \cite{leu_prebiotic_2011}. Similar stalling phenomena appear in other templated assembly processes, including ribosomal translation (where misacylated tRNAs slow peptide bond formation) \cite{Joazeiro2019-wq,Mallory2019-yt,midha_synergy_2023} and, more broadly, in self-assembly systems where defects create kinetic traps \cite{winfree_proofreading_2004,doty_optimizing_2018,lenz_geometrical_2017}, and co-transcriptional ribosomal assembly where checkpoints gate subunit maturation \cite{sanghai_co-transcriptional_2023}.

Per-nucleotide error rates $\mu$ and stalling factors $\eta_\text{stall} \equiv \tau_\text{slow}/\tau_\text{fast} = k_\text{fast}/k_\text{slow}$ vary widely across replication systems. Under what conditions on these parameters would error correction (proofreading) be expected to evolve? Proofreading mechanisms, such as the 3'$\to$5' exonuclease activity of DNA polymerases, function precisely as resets in our framework \cite{hopfield_kinetic_1974,ninio_kinetic_1975,murugan_speed_2012}: they detect stalled trajectories (via the kinetic signature of a mismatch) and restart from the last correct state. The general theory of conditioning developed in Section~\ref{sec:SI-general} provides a quantitative answer. Mapping templated replication onto the two-delta minimal model introduced in this section yields conditions on $\mu$ and $\eta_\text{stall}$ that determine (i) when proofreading increases fitness from speed alone, and (ii) the critical selective value of entropy reduction $s_c$ at which proofreading becomes favorable even when it does not increase speed. See Sec. \ref{subsec:polymer} for the more detailed model of polymerization used in the main text.

\subsection{Two-delta minimal model}

The simplest analytically tractable model of templated replication uses just two completion times: fast (no errors) and slow (one error causes stalling). This binary distribution captures the essential tradeoff with minimal parameters.

\subsubsection{Setup}

The completion-time distribution is:

\begin{equation}
P(T_\text{rep}) = (1 - \mu)\, \delta(T_\text{rep} - \tau_f) + \mu\, \delta(T_\text{rep} - \tau_s),
\label{eq:two-delta}
\end{equation}

with $\tau_s = k\, \tau_f$ and $k > 1$, so that $\tau_f$ ($\equiv \tau_\text{fast}$) and $\tau_s$ ($\equiv \tau_\text{slow}$) represent fast and slow completion times, respectively. The parameter $\mu \in (0,1)$ is the probability of the slow trajectory, the equivalent of error rate. With probability $1-\mu$ the system copies without error and finishes in time $\tau_f$, and with probability $\mu$ it makes an error, stalls, and finishes in time $\tau_s = k\tau_f$.

\subsubsection{Optimal reset time and ordering}

The optimal reset time is $T_r \approx \tau_f$ (just long enough to accept the fast trajectory but reject the slow one). Eq.~\ref{eq:beneficialresets}  applied to the two-delta distribution at $T_r = \tau_f$ --- so that $a = 1 - \mu$, $E(T \mid \text{not lost}) = \tau_f$, and $E(T \mid \text{lost}) = \tau_s$ --- gives the condition for resets to increase speed:

\begin{equation}
\frac{\tau_s}{\tau_f} > 1 + \frac{1}{1 - \mu}.
\label{eq:two-delta-beneficial}
\end{equation}

Eq.~\ref{eq:beneficialresets} should be read as a mean-time proxy for fitness-improving resets. When errors are rare ($\mu \ll 1$), even modest stalling ($\tau_s/\tau_f > 2$ in this proxy) makes resets worthwhile. When errors are common ($\mu$ close to 1), very strong stalling is needed because most attempts will be reset, wasting more time.

The entropy reduction produced at the optimal reset time is the entropy of a binary distribution:

\begin{equation}
\Delta S = -\mu \ln \mu - (1 - \mu) \ln(1 - \mu).
\label{eq:two-delta-DeltaS}
\end{equation}

This is largest at $\mu = 0.5$ (maximum uncertainty about whether an attempt will succeed) and vanishes at $\mu = 0$ or $\mu = 1$ (no uncertainty).

\subsubsection{Distribution of times with resets}

With resets at $T_r$ (where $\tau_f < T_r < k\tau_f$), each replication event consists of some number of failed attempts (each hitting the slow trajectory, probability $\mu$) followed by one fast completion (probability $1-\mu$). This is a geometric retry process:

\begin{equation}
P^\text{reset}(T_\text{rep}) = (1-\mu)\, \delta(T_\text{rep} - \tau_f) + \mu(1-\mu)\, \delta(T_\text{rep} - (T_r + \tau_f)) + \mu^2(1-\mu)\, \delta(T_\text{rep} - (2T_r + \tau_f)) + \cdots
\label{eq:two-delta-P-reset}
\end{equation}

The $n$-th term represents $n$ failed attempts (probability $\mu^n$) each costing $T_r$, followed by one successful fast completion. Summing this geometric series gives the MGF:

\begin{equation}
\tilde{P}^\text{reset}(w) = \frac{(1-\mu)\, e^{w\tau_f}}{1 - \mu\, e^{wT_r}}.
\label{eq:two-delta-MGF}
\end{equation}

\subsubsection{Mapping to templated replication}

The two-delta model maps directly onto templated replication of a sequence of length $L$ with per-step error rate $\mu$:

\begin{itemize}
  \item Correct trajectory ($RR \cdots R$): completion time $\tau_f = L\tau_0$, probability $p_f = (1-\mu)^L \approx 1 - \mu L$.
  \item Erroneous trajectories ($RR \cdots WR \cdots R$, with one mistake $W$): completion time $\tau_s = \eta_\text{stall}\,\tau_f = \eta_\text{stall} L\tau_0$, combined probability $p_s = \mu L$.
\end{itemize}

Here $\tau_0$ is the per-nucleotide incorporation time in the absence of errors and $\eta_\text{stall} = \tau_s/\tau_f$ is the stalling factor, defining how slow is the extension past a mismatch vs a match. The approximation $p_f \approx 1 - \mu L$ is valid when $\mu L \ll 1$, i.e., when most replication events are error-free.

The mean-time proxy condition from Eq.~\ref{eq:two-delta-beneficial}, applied with the substitutions $\mu \to \mu L$ and $\tau_s/\tau_f = \eta_\text{stall}$, becomes

\begin{equation}
\eta_\text{stall} > 1 + \frac{1}{1 - \mu L},
\label{eq:stall-threshold}
\end{equation}

and the order produced by conditioning is

\begin{equation}
\Delta S = -\bigl[(1-\mu L)\ln(1-\mu L) + \mu L\ln(\mu L)\bigr] \simeq \mu L\,[1-\ln(\mu L)],
\label{eq:template-DeltaS}
\end{equation}
this is the completion-time entropy difference, which here equals the path-type entropy difference because each path type --- correct or erroneous --- has a deterministic completion time; cf.\ the discussion in Section~\ref{sec:SI-general}. This connects temporal conditioning (resetting stalled trajectories) to kinetic proofreading: the mechanism exploits the time signature of errors to eliminate them, without directly assessing the microscopic state.

\subsubsection{Critical pressure for proofreading evolution}

We now derive $s_c$, the critical selective value of entropy reduction at which proofreading (conditioning) becomes favorable. This determines, for given $\mu$ and stalling factor $\eta_\text{stall}$, the minimum pressure on trajectory reproducibility needed for error correction to evolve.

Without conditioning, the Euler-Lotka equation gives:

\begin{equation}
2\left[(1-\mu)\, e^{-f_0\, \tau_f} + \mu\, e^{-f_0\, \tau_s}\right] = 1,
\label{eq:template-f0}
\end{equation}

where $\mu$ denotes the per-sequence error probability $\mu_0 L$. This implicit equation determines $f_0$ for given $\mu$, $\tau_f$, and $\tau_s$.

With resets at $T_r$ just above $\tau_f$ (i.e., $\tau_f < T_r < \tau_s$, as in Eq.~\ref{eq:two-delta-P-reset}), only the fast (error-free) trajectories complete before the reset; all erroneous trajectories are destroyed. Taking the limit $T_r \to \tau_f^+$, the Euler-Lotka equation becomes

\begin{equation}
2(1-\mu)\, e^{-f_\text{growth}\, \tau_f} = 1 - \mu\, e^{-f_\text{growth}\, \tau_f},
\label{eq:template-EL-simplified}
\end{equation}

which can be solved directly:

\begin{equation}
e^{-f_\text{growth}\, \tau_f} = \frac{1}{2 - \mu},
\label{eq:template-exp-f}
\end{equation}

giving

\begin{equation}
f_\text{growth} = \frac{\ln(2 - \mu)}{\tau_f}.
\label{eq:f-speed}
\end{equation}

This is one of the few cases where the Euler-Lotka equation has a closed-form solution. The growth rate with resets depends only on the error rate, not on the stalling time --- because the reset eliminates all stalled trajectories before they finish.

The effective fitness with resets is $F = f_\text{growth} \cdot e^{s\Delta S}$, where $\Delta S = -[(1-\mu)\ln(1-\mu)+\mu\ln\mu]$ is the binary Shannon entropy difference for the two-delta model. Proofreading becomes favorable when $F > f_0$. Taking logarithms, the critical pressure is

\begin{equation}
s_c = \frac{\ln(f_0 / f_\text{growth})}{\Delta S} = \frac{\ln(f_0 / f_\text{growth})}{-[(1-\mu)\ln(1-\mu)+\mu\ln\mu]}.
\label{eq:sc-formula}
\end{equation}

The numerator is the growth-rate cost of resets (how much slower the population grows due to discarding erroneous attempts); the denominator is the entropy-reduction benefit (how much more reproducible the surviving trajectories are). When $f_\text{growth} \geq f_0$ (resets speed up growth), the numerator is zero or negative, giving $s_c = 0$: proofreading pays for itself through speed alone.

\subsection{Reference values for biological systems}
\label{sec:SI-bio-references}

The biological reference points overlaid on Fig.~\ref{fig:poly}(b) of the main text consist of per-nucleotide error rates $\mu_0$ and per-mismatch stalling factors $\eta_\text{stall}$ for five representative templated-replication systems: non-enzymatic DNA and RNA copying, a ribozyme, and two DNA polymerases (T7 and Taq). For each system, multiple literature values are used for both $\mu_0$ and $\eta_\text{stall}$; the marker location shows the median and the error bars span the interquartile range (or the min--max range when fewer than four values are available). Table~\ref{tab:mu_sources} lists every $\mu_0$ value plotted, together with its primary source. The corresponding stalling-factor tables are taken from the compilation in \cite{ravasio_evolution_2026} (Supplementary Tables~S1.3 and~S1.5 therein), to which the reader is referred for the original per-mismatch references.

\begin{table*}[h]
\centering
\begin{tabular}{l l l}
\hline
System & $\mu_0$ & Source \\
\hline
Non-enz.\ DNA           & $7.6\times10^{-2}$     & \cite{rajamani_effect_2010} \\
Non-enz.\ RNA           & $2\times10^{-2}$       & \cite{prywes_nonenzymatic_2016} \\
Non-enz.\ RNA           & $1.7\times10^{-1}$     & \cite{leu_prebiotic_2011} \\
Ribozyme                & $10^{-2}$              & \cite{johnston_rna-catalyzed_2001} \\
T7 DNA pol.\ (exo$^-$, Sequenase)  & $3.4\times10^{-5}$     & \cite{keohavong_fidelity_1989} \\
T7 DNA pol.\ (exo$^-$, Sequenase)  & $4.4\times10^{-5}$     & \cite{ling_optimization_1991,cariello_fidelity_1991} \\
T7 DNA pol.\ (exo$^-$, Sequenase)  & $2\times10^{-4}$       & \cite{eckert_dna_1991} \\
Taq DNA pol.            & $10^{-5}$              & \cite{eckert_dna_1991} \\
Taq DNA pol.            & $8.9\times10^{-5}$     & \cite{cariello_fidelity_1991} \\
Taq DNA pol.            & $1.1\times10^{-4}$     & \cite{tindall_fidelity_1988} \\
Taq DNA pol.            & $2\times10^{-4}$       & \cite{eckert_dna_1991} \\
$\phi29$ DNA pol.\ (exo$^-$, D12A--THR)
                        & $1.5\times10^{-4}$     & \cite{handal_marquez_directed_2022} \\
$\phi29$ DNA pol.\ (exo$^-$, D12A/D66A; inferred)\footnotemark
                        & $2.3\times10^{-4}$     & \cite{paez_genome_2004,perez_arnaiz_involvement_2006} \\
\hline
\end{tabular}
\caption{Per-nucleotide error rates $\mu_0$ used as biological reference points in Fig.~\ref{fig:poly}(b). Each plotted array element is listed individually with its primary source.}
\label{tab:mu_sources}
\footnotetext{The $\phi29$ D12A/D66A exo$^-$ value is inferred by multiplying the wild-type $\phi29$ error rate measured by Paez et al. during whole genome amplification ($9.5\times10^{-6}$ errors/bp) by the 24-fold reduction in fidelity reported for the D12A/D66A exonuclease-deficient mutant by P{\'e}rez-Arnaiz et al., giving $24 \times 9.5\times10^{-6} \approx 2.3\times10^{-4}$ errors/bp.}
\end{table*}

\subsection{Stochastic kinetic model of templated replication}
\label{subsec:polymer}

The two-delta model captures the essential trade-offs analytically but abstracts away the microscopic kinetics. Here we describe the explicit stochastic model used to generate main-text Fig.~\ref{fig:poly}, Fig.~\ref{fig:SI_stalling_exo}, and~Fig.~\ref{fig:SI_exo_vs_full}. This model tracks polymerisation monomer by monomer, incorporates stalling after misincorporation, and admits two distinct mechanisms of conditioning: exonuclease activity (partial reset, removing only the last monomer) and full-chain degradation (full reset, destroying the entire chain). For the no-conditioning and full-reset mechanisms we derive exact Laplace transforms of the completion-time distribution, enabling growth-rate calculations without Monte Carlo sampling. For exonuclease-mediated partial reset we derive an approximate form.

\subsubsection{Kinetic scheme}

A template of length $L$ is copied sequentially. The state of the system is the current chain length $n \in \{0, 1, \ldots, L\}$ and the identity (correct or incorrect) of each incorporated monomer. The kinetic rates are:

\begin{itemize}
  \item \emph{First position} ($n = 0$): the monomer is added at rate $k_\text{fast}$ (no stalling possible, since there is no preceding monomer to mismatch).
  \item \emph{Subsequent positions} ($n \geq 1$): if the preceding monomer is correct, the next addition occurs at rate $k_\text{fast}$; if incorrect, the addition rate drops to $k_\text{slow} = k_\text{fast} / \eta_\text{stall}$, where $\eta_\text{stall} \geq 1$ is the dimensionless stalling factor.
  \item \emph{Error probability}: each newly added monomer is correct with probability $1 - \mu_0$ and incorrect with probability $\mu_0$, independent of position and history. The stalling factor $\eta_\text{stall} = k_\text{fast}/k_\text{slow}$ is the same dimensionless quantity used in the socks-before-shoes section; for exponential waiting times it equals the time-cost ratio $\tau_\text{slow}/\tau_\text{fast}$.
  \item \emph{Conditioning}: at each position $n \geq 1$, a resetting event competes with addition at rate $k_r$. Two mechanisms are considered: exonuclease activity (removes only the last monomer, $n \to n-1$) and full reset (destroys the entire chain, $n \to 0$).
\end{itemize}

The system resets at rate $k_r$ regardless of monomer identity and selectivity arises from stalling: wrong monomers slow the forward rate to $k_\text{slow} \ll k_\text{fast}$, so the correction has $\eta_\text{stall}$ times longer to fire before the next monomer is added. This is the same temporal-conditioning principle as the rest of the paper: the mechanism does not recognize the error directly, but exploits the time signature of errors to preferentially remove them.

At each position, the added monomer is either correct (probability $1-\mu_0$, forward rate $k_\text{fast}$) or incorrect (probability $\mu_0$, forward rate $k_\text{slow}$). The monomer survives if addition wins, with probability $k_\text{add}/(k_\text{add} + k_r)$ where $k_\text{add}$ is the forward rate at the next position. Averaging over monomer identity, the fraction of surviving monomers that are incorrect gives the effective per-site error rate:
\begin{equation}
\mu_\text{eff} = \frac{\mu_0\, k_\text{slow} / (k_\text{slow} + k_r)}{(1-\mu_0)\, k_\text{fast} / (k_\text{fast} + k_r) + \mu_0\, k_\text{slow} / (k_\text{slow} + k_r)},
\label{eq:poly-mu-eff}
\end{equation}
which is the bulk per-site error rate for positions that must survive until the next incorporation event. Strictly, the final incorporated monomer is a boundary site: replication stops immediately when $n=L$, so the last monomer is not exposed to a subsequent addition-versus-correction race and has error probability $\mu_0$. For a finite chain, the corresponding sequenced error rate is therefore
\begin{equation}
\epsilon_L \simeq \frac{L-1}{L}\,\mu_\text{eff} + \frac{1}{L}\,\mu_0
= \mu_\text{eff} + \frac{\mu_0-\mu_\text{eff}}{L}.
\label{eq:poly-mu-eff-finiteL}
\end{equation}
For the simulations in Fig.~\ref{fig:SI_stalling_exo}, $L=10^5$, so this boundary correction is negligible; the plotted black lines use the bulk expression in Eq.~\ref{eq:poly-mu-eff}.

Replication completes when $n = L$. In the Gillespie simulation, at each step the waiting time is drawn from an exponential distribution with rate equal to the sum of all active rates, and the transition is chosen proportionally. The Gillespie algorithm is a standard method for exact stochastic simulation of chemical reaction networks: it generates statistically exact trajectories of the master equation without any approximation beyond the use of random numbers. We use it here as a numerical check on the deterministic calculations described below: Fig.~\ref{fig:SI_stalling_exo} checks the bulk effective error rate $\mu_\text{eff}$, with the finite-$L$ boundary correction in Eq.~\ref{eq:poly-mu-eff-finiteL} negligible for the plotted $L=10^5$, and Fig.~\ref{fig:SI_validation} checks the exact full-reset growth-rate transform and the approximate exonuclease growth-rate transform.

\subsubsection{Why we use Laplace transforms}

The Laplace transform is the natural mathematical tool for this problem because the Euler-Lotka equation itself \emph{is} a Laplace transform: solving $2\hat{\phi}(f_\text{growth}) = 1$ for the growth rate $f_\text{growth}$ requires evaluating the Laplace transform $\hat{\phi}(\lambda) = \int_0^\infty e^{-\lambda T} P(T)\, dT$ of the completion-time distribution at $\lambda = f_\text{growth}$. If we can compute or approximate $\hat{\phi}(\lambda)$ directly, we can find the growth rate by root-finding without sampling completion times or constructing histograms. The no-correction and full-reset cases are exact and free of statistical noise; the exonuclease case uses the deterministic approximation described below.

\subsubsection{Per-step building block}

At each position $n \geq 1$, addition and correction are independent Poisson processes; whichever fires first determines whether the chain advances or is corrected. Averaging over the identity of the preceding monomer (correct with probability $1-\mu_0$, wrong with probability $\mu_0$) gives the per-step Laplace-weighted probabilities:
\begin{align}
\alpha(\lambda) &= \frac{(1-\mu_0)\, k_\text{fast}}{k_\text{fast} + k_r + \lambda} + \frac{\mu_0\, k_\text{slow}}{k_\text{slow} + k_r + \lambda}, \label{eq:poly-alpha} \\
\beta(\lambda) &= \frac{(1-\mu_0)\, k_r}{k_\text{fast} + k_r + \lambda} + \frac{\mu_0\, k_r}{k_\text{slow} + k_r + \lambda}, \label{eq:poly-beta}
\end{align}
where $\alpha$ is the Laplace-weighted probability of advancing (adding a monomer) and $\beta$ is the Laplace-weighted probability of a correction event. At $\lambda = 0$, direct evaluation gives $\alpha(0) + \beta(0) = 1$ (one of the two events fires with probability one). For $\lambda > 0$, $\alpha(\lambda) + \beta(\lambda) = E[e^{-\lambda\tau}] < 1$, where $\tau$ is the waiting time for whichever event fires first; the deficit $1 - \alpha(\lambda) - \beta(\lambda) = 1 - E[e^{-\lambda\tau}]$ is the standard time discount of the Laplace transform, capturing the cost of waiting before either event fires. This is why $\alpha(\lambda)$ and $\beta(\lambda)$ are called \emph{Laplace-weighted} probabilities rather than raw probabilities. At position $n = 0$ there is no preceding monomer and no correction (nothing to correct), so the step is a simple exponential wait: $\psi_0(\lambda) = k_\text{fast}/(k_\text{fast} + \lambda)$.

\begin{figure}
  \includegraphics[width=0.8\columnwidth]{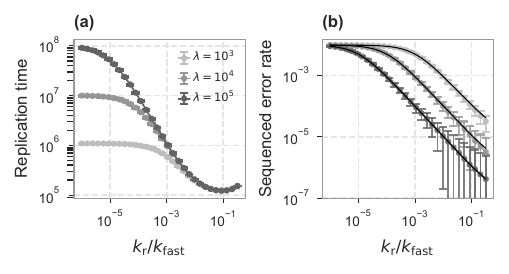}
  \caption{Gillespie simulation of exonuclease-mediated error correction. A polymer of target length $L = 10^5$ is synthesized with per-step error rate $\mu_0 = 10^{-2}$ and three stalling factors $\eta_\text{stall} = k_\text{fast}/k_\text{slow}$ ($N = 5{,}000$ replicates per point; error bars show $\pm 1$ s.d.). (a) Mean replication time $\langle T \rangle$ (total Gillespie clock time to reach $n = L$) versus correction rate $k_r/k_\text{fast}$. Increasing $k_r$ initially reduces time by excising stalled positions, but excessive $k_r$ erases correct monomers and slows completion. The optimal $k_r$ balances these effects. (b) Sequenced error rate (fraction of incorrect monomers in the completed product) versus $k_r$. For finite $L$, this differs from the bulk expression $\mu_\text{eff}$ only by the final-site boundary correction in Eq.~\ref{eq:poly-mu-eff-finiteL}, which is negligible here. Higher stalling factors yield steeper error-rate reduction because mismatched positions are exposed to the exonuclease for longer. Black lines: bulk prediction from Eq.~\ref{eq:poly-mu-eff}.}
  \label{fig:SI_stalling_exo}
\end{figure}

\subsubsection{Laplace transform: no correction (baseline)}

Without any correction mechanism ($k_r = 0$), all $L$ steps are independent: each monomer is added without any possibility of removal. The total completion time is the sum of $L$ independent waiting times. The Laplace transform of a sum of independent variables is the product of their individual transforms, giving
\begin{equation}
\hat{\phi}_0(\lambda) = \psi_0(\lambda) \cdot \alpha_0(\lambda)^{L-1} = \frac{k_\text{fast}}{k_\text{fast} + \lambda} \left[\frac{(1-\mu_0)\, k_\text{fast}}{k_\text{fast} + \lambda} + \frac{\mu_0\, k_\text{slow}}{k_\text{slow} + \lambda}\right]^{L-1}.
\label{eq:poly-phi0}
\end{equation}

\subsubsection{Laplace transform: full reset}

When the correction fires at any position $n \geq 1$, the entire chain is discarded and synthesis restarts from $n = 0$. This has the structure of a geometric retry: each attempt either succeeds (the chain reaches length $L$ without any correction firing) or fails (a correction fires at some position, triggering a restart). After failure, the system faces the same problem from scratch.

A single attempt either succeeds (Laplace factor $\hat{\phi}_\text{succ} = \psi_0\, \alpha^{L-1}$) or fails at some position $n$. Summing over all failure positions gives $\hat{\phi}_\text{fail} = \psi_0\, \beta\, (1 - \alpha^{L-1})/(1 - \alpha)$. The self-consistency equation $\hat{\phi} = \hat{\phi}_\text{succ} + \hat{\phi}_\text{fail} \cdot \hat{\phi}$ (success, or failure followed by the whole problem again) gives

\begin{equation}
\hat{\phi}_\text{full}(\lambda) = \frac{\hat{\phi}_\text{succ}(\lambda)}{1 - \hat{\phi}_\text{fail}(\lambda)}.
\label{eq:poly-phi-full}
\end{equation}

This is analogous to repeatedly flipping a biased coin until heads: the Laplace transform is a ratio involving the probability of success on each trial.

\subsubsection{Laplace transform: exonuclease (partial reset)}

With exonuclease activity, a correction event at position $n$ removes only the last monomer ($n \to n-1$), preserving the previously copied prefix. This is the polymer analog of a save point: progress is not fully lost.

Unlike full resets, this couples neighboring positions. The time to advance past position $n$ depends on what happens if the exonuclease chews back: the chain drops back to position $n-1$, and the system must re-advance through $n-1$ before making another attempt at $n$. The time spent at position $n$ therefore depends on the time needed at position $n-1$, which in turn depends on $n-2$, and so on. These times depend on the sequence of the polymer, as a polymer with more incorrect monomers is likely to take longer. We reduce this complexity by averaging over the identity of monomers.

Consider a polymer that currently has length $n$ with sequence $\sigma \in \{r, w\}^n$, where $r$ denotes the right monomer and $w$ denotes the wrong monomer (i.e. a mutation). Denote by $T_{n,\sigma}$ the time taken for this polymer to reach the terminal length $L$. We now imagine that one of two things happen:

\begin{itemize}

\item The exonuclease chews back with rate $k_r$ -- leading to a polymer of length $n-1$ and sequence $\sigma^-$ (i.e., $\sigma$ with the last monomer removed).

\item Polymerization has a rate $k(\sigma)$. Note that this rate is sequence dependent -- $k(\sigma) = k_{\text{fast}}$ if the last monomer of $\sigma$ is $r$, and $k(\sigma) = k_{\text{slow}}$ if the last monomer of $\sigma$ is $w$.

This gives a polymer of length $n+1$, and a new sequence. With probability $1-\mu_0$, this new sequence will be $\sigma$ appended with the right monomer (which we will denote $\sigma + r$), and with probability $\mu_0$ it will be $\sigma$ appended with the wrong monomer, $\sigma + w$.

\end{itemize}

The new completion time (i.e. the time remaining for the polymer to reach length $L$) depends on the old completion time $T_{n,\sigma}$ via the recursion relationship,
\[ T_{n,\sigma} = \tau + T_{\text{next state}}, \]
where $\tau$ is the time spent waiting for either the exonuclease or polymerase to fire. As before, we will be interested in the Laplace transform,
\[ \phi_{n,\sigma} \equiv \langle \exp \left( -\lambda T_{n,\sigma} \right) \rangle = \langle e^{-\lambda \tau} \exp \left( - \lambda T_{\text{next state}} \right) \rangle. \]
We average over the next step: i.e., whether the exonuclease chewed back (with probability $k_r/(k_r + k_{\sigma})$) or polymerization happened (with probability $k_{\sigma}/(k_r + k_{\sigma})$),
\[ \phi_{n,\sigma} = \frac{k(\sigma)}{k(\sigma) + k_r} \frac{k(\sigma) + k_r}{\lambda + k(\sigma) + k_r} \phi_{+} (\lambda) + \frac{k_r}{k(\sigma) + k_r} \frac{k(\sigma) + k_r}{\lambda + k(\sigma) + k_r} \phi_{-} (\lambda) = \frac{1}{\lambda + k(\sigma) + k_r} \left(k(\sigma) \phi_+(\lambda) + k_r \phi_-(\lambda) \right), \]
where
\[ \phi_+(\lambda) \equiv (1-\mu_0) \, \phi_{n+1, \sigma + r}(\lambda) + \mu_0 \, \phi_{n+1,\sigma + w}(\lambda), \]
\[ \phi_-(\lambda) \equiv \phi_{n-1,\sigma^-} (\lambda). \]
At this point, everything still depends on the entire sequence $\sigma$ and the calculation is intractable. We now make a key approximation: namely, that we only keep track of the identity of the last base,
\[ \phi_{n, \sigma} \approx \phi_{n, r} \text{ or } \phi_{n, w}, \]
where the $r$ and $w$ in the subscripts denote that the last base is either $r$ or $w$, respectively. This is approximate in that, when the exonuclease acts and reveals the preceding base, we have to assume that it is $r$ with probability $1-\mu_0$ and $w$ with probability $\mu_0$:
\[ \phi_-(\lambda) \approx (1-\mu_0)\phi_{n-1, r} + \mu_0 \phi_{n-1, w}, \]
We then define a new quantity, $\phi_n(\lambda)$, which averages over the identity of the last base,
\[ \phi_n(\lambda) \equiv (1-\mu_0)\phi_{n, r} + \mu_0 \phi_{n, w}. \]
In terms of this new quantity, $\phi_+(\lambda) = \phi_{n+1}(\lambda)$, $\phi_-(\lambda) \approx \phi_{n-1}(\lambda)$, and the recursion relation becomes (after some algebra)
\[ \phi_n = \alpha(\lambda) \, \phi_{n+1}(\lambda) + \beta(\lambda) \, \phi_{n-1}(\lambda), \]
with $\alpha(\lambda)$ and $\beta(\lambda)$ as defined in the last section. To solve the recursion, we divide by $\phi_{n+1}$ and introduce the ratio
\[ \psi_n(\lambda) \equiv \frac{\phi_n}{\phi_{n+1}}, \]
to obtain a partial fraction representation valid for $n \geq 1$:
\[ \psi_n = \frac{\alpha(\lambda)}{1 - \beta(\lambda) \psi_{n-1}}. \]
When $n=0$, the exonuclease cannot act and so
\[ \phi_{0}(\lambda) = \frac{k_{\text{fast}}}{\lambda + k_{\text{fast}}} \, \phi_1(\lambda) \Rightarrow \psi_0(\lambda) = \frac{k_{\text{fast}}}{\lambda + k_{\text{fast}}}. \]
Now, we are interested in the completion time starting from $n=0$. From the partial fraction recursion and the initial condition of $\psi_{0}(\lambda)$, $\psi_n(\lambda)$ can be calculated numerically at each value of $\lambda$. To recover $\phi_0(\lambda)$ -- which we also denote $\phi_{\text{exo}}(\lambda)$ to compare against $\phi_{\text{full}}(\lambda)$, we use the definition of $\psi_n$,
\begin{equation}
    \label{eq:poly-phi-exo}
    \phi_{\text{exo}}(\lambda) = \frac{\phi_0}{\phi_1} \frac{\phi_1}{\phi_2} \cdots \frac{\phi_{L-1}}{\phi_L} \times \phi_L =  \prod_{n=0}^{L-1} \psi_n(\lambda)
\end{equation}
where we have used the fact that $\phi_L(\lambda) = 1$ by definition.

\begin{figure}
  \includegraphics[width=0.6\columnwidth]{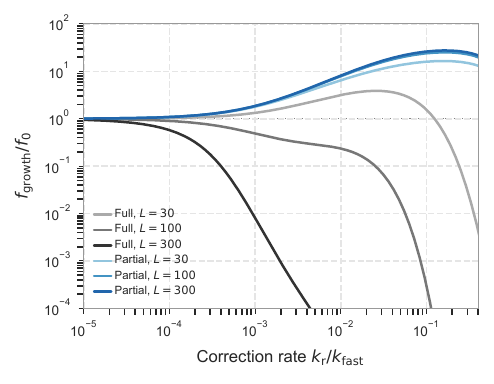}
  \caption{Exonuclease (partial reset) versus full reset in the polymerisation model. Growth-rate fold-change $f_\text{growth}/f_0$ versus correction rate $k_r/k_\text{fast}$ for $\eta_\text{stall} = 10^3$, $\mu_0 = 0.05$, and three polymer lengths $L \in \{30, 100, 300\}$. Gray: full resets (entire chain discarded). Blue: exonuclease (last monomer removed). The horizontal dotted line marks $f_\text{growth}/f_0 = 1$ (no benefit). Full resets improve growth rate only in a narrow rate window that shrinks with $L$; at high rates they become catastrophic because each reset discards the entire $O(L)$-length prefix. Exonuclease maintains a broad beneficial window and saturates at a plateau that increases with $L$, because the save-point mechanism preserves the previously copied prefix. This directly parallels the save-point results in the socks-before-shoes model. Full-reset curves use the exact transform in Eq.~\eqref{eq:poly-phi-full}; exonuclease curves use the length-only approximation in Eq.~\eqref{eq:poly-phi-exo}. Both are evaluated on 200 log-spaced rate points.}
  \label{fig:SI_exo_vs_full}
\end{figure}

\begin{figure}
  \includegraphics[width=\columnwidth]{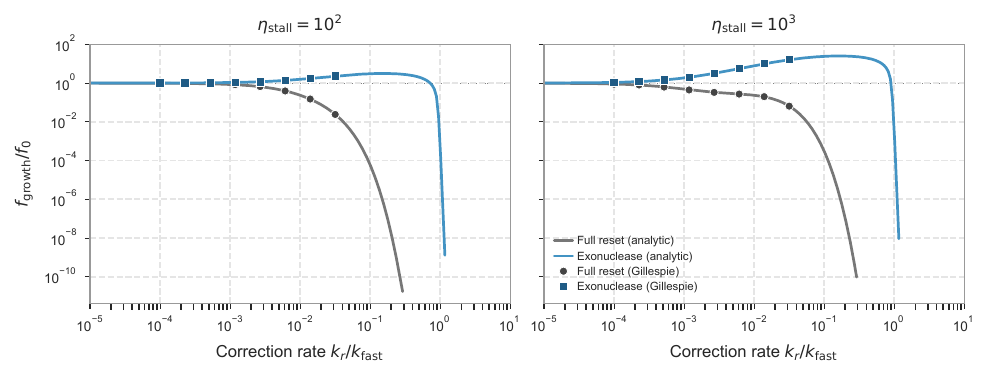}
  \caption{Validation of Laplace-transform growth-rate calculations against Gillespie simulation. Growth-rate fold-change $f_\text{growth}/f_0$ versus correction rate $k_r/k_\text{fast}$ for $L = 100$, $\mu_0 = 0.05$, at two stalling factors $\eta_\text{stall} \in \{10^2, 10^3\}$ (panels). Solid lines: exact full-reset transform (gray, Eq.~\eqref{eq:poly-phi-full}) and length-only exonuclease approximation (blue, Eq.~\eqref{eq:poly-phi-exo}), evaluated at 200 log-spaced rate points. Markers: Gillespie Monte Carlo estimates of $f_\text{growth}$ obtained by solving $2\langle e^{-f_\text{growth}\,T}\rangle = 1$ on $N = 5{,}000$ sampled completion times per rate ($\bigcirc$ full reset, $\square$ exonuclease, 8 log-spaced rate points). Individual validation trajectories are simulated with a maximum clock time $T_\text{max}=10^8$; the Monte Carlo rate grid is therefore restricted to $k_r \leq 10^{-1.5}$, avoiding the high-rate full-reset collapse regime where completion times become strongly censored. The full-reset transform agrees with MC within sampling uncertainty, while the exonuclease approximation tracks the Gillespie estimates to within a few percent over the tested range, including the regime where exonuclease saturates above $1$ and full reset collapses below $1$. The exonuclease curve is computed via the continued-fraction recursion in Eq.~\eqref{eq:poly-phi-exo}.}
  \label{fig:SI_validation}
\end{figure}

\subsubsection{Growth rate via the Euler-Lotka equation}

For each mechanism (baseline, full reset, exonuclease), the population growth rate $f_\text{growth}$ is determined by
\begin{equation}
2\,\hat{\phi}(f_\text{growth}) = 1,
\label{eq:poly-euler-lotka}
\end{equation}
where $\hat{\phi}$ is the exact transform from Eq.~\eqref{eq:poly-phi0} or Eq.~\eqref{eq:poly-phi-full}, or the approximate length-only transform from Eq.~\eqref{eq:poly-phi-exo}. This equation is solved numerically via Brent's method. For the correction mechanisms, the growth rate is maximized over $k_r$, yielding the optimal fold-change $\max_{k_r} f_\text{growth} / f_0$.

\subsubsection{Simulation parameters and figures}

The main-text figure (Fig.~\ref{fig:poly}) shows $\max_{k_r} f_\text{growth}/f_0$ versus polymer length $L$ for two error rates: $\mu_0 = 5 \times 10^{-2}$ (error-prone, prebiotically relevant) and $\mu_0 = 10^{-5}$ (accurate, enzymatic). For each, six stalling factors are compared: $\eta_\text{stall} \in \{1, 10, 10^2, 10^3, 10^4, 10^6\}$. Solid lines show exonuclease (partial reset); dashed lines show full reset. The correction rate is optimized over a grid of 80 log-spaced values in $[10^{-6}, 10]$ (in units of $k_\text{fast}$). Polymer lengths range from $L = 3$ to $L = 3 \times 10^4$ ($\mu_0 = 5 \times 10^{-2}$) or $L = 10^7$ ($\mu_0 = 10^{-5}$, for $\eta_\text{stall} \in \{10^4, 10^6\}$).

Fig.~\ref{fig:SI_stalling_exo} compares the local bulk error-rate prediction with Gillespie simulations ($N = 5{,}000$ replicates, $L = 10^5$, $\mu_0 = 10^{-2}$): panel~(a) shows Gillespie-sampled mean replication times, and panel~(b) overlays the bulk $\mu_\text{eff}$ prediction (Eq.~\ref{eq:poly-mu-eff}, black lines) on the sequenced error rate. As the correction rate $k_r$ increases, replication time decreases (stalled positions are excised) and the sequenced error rate drops (wrong monomers are preferentially removed). At excessive $k_r$, correct monomers are also removed, and the time begins to increase again; the error rate continues to decrease monotonically because the selectivity ratio $k_\text{fast}/(k_\text{fast} + k_r)$ vs.\ $k_\text{slow}/(k_\text{slow} + k_r)$ always favors removing wrong monomers (Eq.~\ref{eq:poly-mu-eff}).

Fig.~\ref{fig:SI_exo_vs_full} is the equivalent of the savepoints figure of the socks-before-shoes model shown in Fig. \ref{fig:savepoints} in the main.

\putbib
\end{bibunit}

\end{document}